\documentclass[10pt,journal,compsoc]{IEEEtran}

\usepackage[nocompress]{cite}
\usepackage{times}
\usepackage{url}
\usepackage{multirow}
\usepackage{verbatim}
\usepackage{color}
\usepackage[dvipsnames]{xcolor}
\usepackage{graphicx}

\def\toperror{\tau}
\def\angular{\theta}
\def\tsdist{\delta}

\begin{document}

\title{Evaluating Cartogram Effectiveness}
%
%
%
%

\author{Sabrina~Nusrat,
	Md. Jawaherul Alam,
	Stephen Kobourov
\IEEEcompsocitemizethanks{\IEEEcompsocthanksitem S.~Nusrat and S.~Kobourov are at the University of Arizona, Tucson, USA. \protect\\
E-mail: \{sabrinanusrat,kobourov\}@cs.arizona.edu
\IEEEcompsocthanksitem Md. Jawaherul Alam is at the University of California, Irvine, USA. \protect\\
E-mail: alamm1@uci.edu}
}

\markboth{Evaluating Cartogram Effectiveness}%
{{Nusrat, Alam and Kobourov}}
%


\IEEEtitleabstractindextext{%
\begin{abstract}
Cartograms are maps in which areas of geographic regions, such as countries and states, appear in proportion to some variable of interest, such as population or income.
Cartograms are popular visualizations for geo-referenced data that have been used for over a century to illustrate patterns and trends in the world around us. Despite the popularity of cartograms, and the large number of cartogram types, there are few studies evaluating the effectiveness of cartograms in conveying information. Based on a recent task taxonomy for cartograms, we evaluate four major types of cartograms: contiguous, non-contiguous, rectangular, and Dorling cartograms. We first evaluate the effectiveness of these cartogram types by quantitative performance analysis (time and error). Second, we collect qualitative data with an attitude study and by analyzing subjective preferences. Third, we compare the quantitative and qualitative results with the results of a metrics-based cartogram evaluation. Fourth, we analyze the results of our study in the context of cartography, geography, visual perception, and demography.
Finally, we consider implications for design and possible improvements.
\end{abstract}

}

\maketitle

\IEEEdisplaynontitleabstractindextext
\IEEEpeerreviewmaketitle

\section{Introduction}

Cartograms are maps in which areas of geographic regions, such as countries and states, appear in proportion to some variable of interest, such as population or income. They are popular visualizations for geo-referenced data that have been used for over a century~\cite{Tobler04}. As such visualizations make it possible to gain insight into patterns and
trends in the world around us, they have gained a great deal of attention from researchers in computational cartography, geography, computational geometry, and GIS.  Many different types of cartograms have been proposed and implemented, optimizing different aspects: statistical accuracy (cartographic error), geographic accuracy (preserving the outlines of geographic shapes), and topological accuracy (maintaining correct adjacencies between countries).

\begin{figure}
{
\centering
\includegraphics[width=0.23\textwidth]{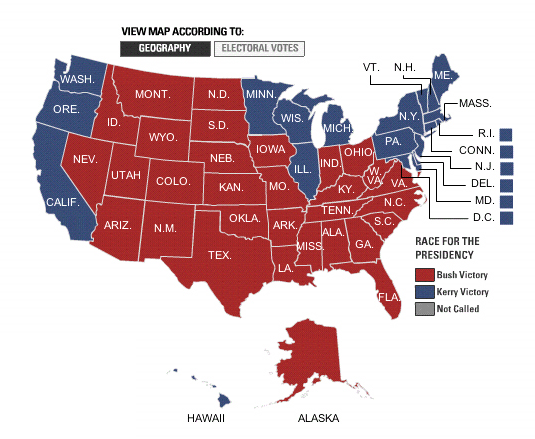} 
\includegraphics[width=0.23\textwidth]{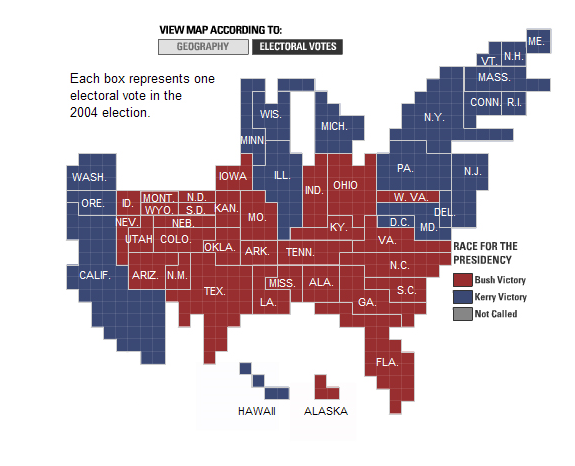}
\caption{Geographic map and a cartogram for the 2004 US election~\cite{NYT_04}.}
\label{fig:red-blue}
}
\end{figure}

Cartograms provide a compact and visually appealing way to represent the world's political, social and economic state in pictures. Red-and-blue population cartograms of the United States have become an accepted standard for showing political election predictions and results.
Likely due to aesthetic appeal and the possibility to 
put political and socioeconomic data into perspective, cartograms are widely used  in newspapers, magazines, textbooks, and blogs. 
For example, while geographically  accurate maps seemed to show an overwhelming victory for George~W.~Bush in the 2004 election; the population cartograms used by the New York Times~\cite{NYT_04}
 effectively communicated the near even split; see Fig.~\ref{fig:red-blue}. 
The Los Angeles Times~\cite{LAT12} shows the 2012 election results using cartograms and cartograms are used to show the European Union election results of 2009 in the Dutch daily newspaper NRC~\cite{NRC}. In addition to visualizing elections, cartograms are frequently used to represent other kinds of geo-referenced data. Dorling cartograms are used in the UK Guardian newspaper~\cite{Guar} to visualize social structure and in the New York Times to show the distribution of medals in Olympic Games since 2008~\cite{NYT16}. 
Popular TED talks use cartograms to illustrate how the news media can present a distorted view of the world~\cite{Alisa}, 
and to illustrate the progress of developing countries~\cite{Hans2}.
 Cartograms continue to be used in textbooks, for example, to teach middle-school and high-school students about global demographics and human development~\cite{Class2}.

Despite the popularity of cartograms and the large number of cartogram variants, there are very few studies evaluating  cartograms. In order to design effective visualizations we need to compare cartograms generated by different methods on a variety of suitable tasks. 
In this paper we describe an in-depth evaluation of four major types of cartograms: contiguous, non-contiguous, rectangular, and Dorling cartograms. We first evaluate the effectiveness of these cartogram types by quantitative performance analysis (time and error) with a controlled experiment that covers seven different tasks from a recently developed task taxonomy for cartograms~\cite{Task_C}. 
Second, we collect qualitative data with an attitude study and by analyzing subjective preferences.
Third, we compare the quantitative and qualitative results with the results of a metrics-based cartogram evaluation. Fourth, we analyze the results of our study in the context of cartography, geography, visual perception, and demography. Finally, we consider implications for design and possible improvements.


\section{Related Work}


Cartograms have a  long history; several major types of cartograms are briefly reviewed  in Sec.~\ref{sec:types}.
While there is some work on quantitative comparisons between the different types, there is no systematic qualitative evaluation. 

In 1975 Dent~\cite{dent1975} considered the effectiveness of cartograms and wrote that ``attitudes point out that these (value-by-area) cartograms are thought to be confusing and difficult to read; at the same time they appear interesting, generalized, innovative, unusual, and having -- as opposed to lacking -- style''. Dent also suggested effective communication strategies when the audience is not familiar with the underlying geography and statistics, e.g., providing an inset map and labeling the statistical units on the cartogram.
Griffin~\cite{Gri83} studied the task of identifying locations in cartograms and found that cartograms are effective. Olson~\cite{Olson} designed methods for the construction of non-contiguous cartograms and studied their characteristics. 
Krauss~\cite{Krauss_ms} also studied non-contiguous cartograms
using three evaluation tasks (from very general to specific) in order to find out how well the geographic information is communicated and concluded that non-contiguous cartograms work well for showing general distribution, but not for showing specific information (e.g., ratios between two regions). 
Sun and Li~\cite{Hui} analyzed the effectiveness of different types of maps by collecting subjective preferences. Two types of experimental tests were conducted: (1) comparison of cartograms with thematic maps (choropleth maps, proportional symbol maps and dot maps), and (2) comparison between cartograms (non-contiguous cartogram, diffusion cartogram, rubber sheet cartogram, Dorling cartogram, and pseudo-cartogram). 
The participants in this study were  asked to select one map that is more effective for the representation of the given dataset and to provide reasons for this choice. 
The results indicate that cartograms are more effective in the representation of nominal data 
(e.g., who who won--republicans or democrats?), but thematic maps are more effective in the representation ordinal data (e.g., population growth rates). Note that in both experiments the subjects gave their preferences, but were not asked to perform any specific tasks.

In a more recent study, Kaspar et al.~\cite{kaspar2013empirical} investigated how people make sense of population data depicted in contiguous (diffusion) cartograms, compared to choropleth maps, augmented with graduated circle maps. The subjects were asked to perform tasks, based on Bertin's map reading levels (\textit{elementary}, \textit{intermediate} and \textit{overall})~\cite{BERTIN83}. The overall results showed that the augmented choropleth maps are more effective (as measured by accurate responses) and more efficient (as measured by faster responses) than the cartograms.
The results seemed to depend on the complexity of the task (simple tasks are easier to perform in both maps compared to complex tasks), and the shape of the polygons. Note that only one type of cartogram (Gastner-Newman diffusion~\cite{GN04}) was used in this study.

In order to improve cartogram design, Tao~\cite{Manting} conducted an online survey to collect suggestions from map users. 
The majority of the participants found cartograms difficult to understand 
but at least agreed that 
 cartograms are commonly regarded as members of the map ``family''. 
  Jennifer Ware~\cite{Jen} evaluated the effectiveness of animation in cartograms with a user-study in which \textit{locate} and \textit{compare} tasks were considered. The results indicate that although the participants preferred animated cartograms, the response time for the tasks was best in static cartograms. 

The studies above indicate an interest in cartograms and their effectiveness. While some specific types of cartograms have been evaluated on some specific tasks, a more comprehensive evaluation of different types of cartograms with a varied set of questions is lacking. In this paper we consider both qualitative and quantitative measurements, covering the spectrum of cartogram tasks, using four of the main types of cartograms.

\begin{figure*}[t!]
\centering
\hfill
\includegraphics[width=0.48\textwidth]{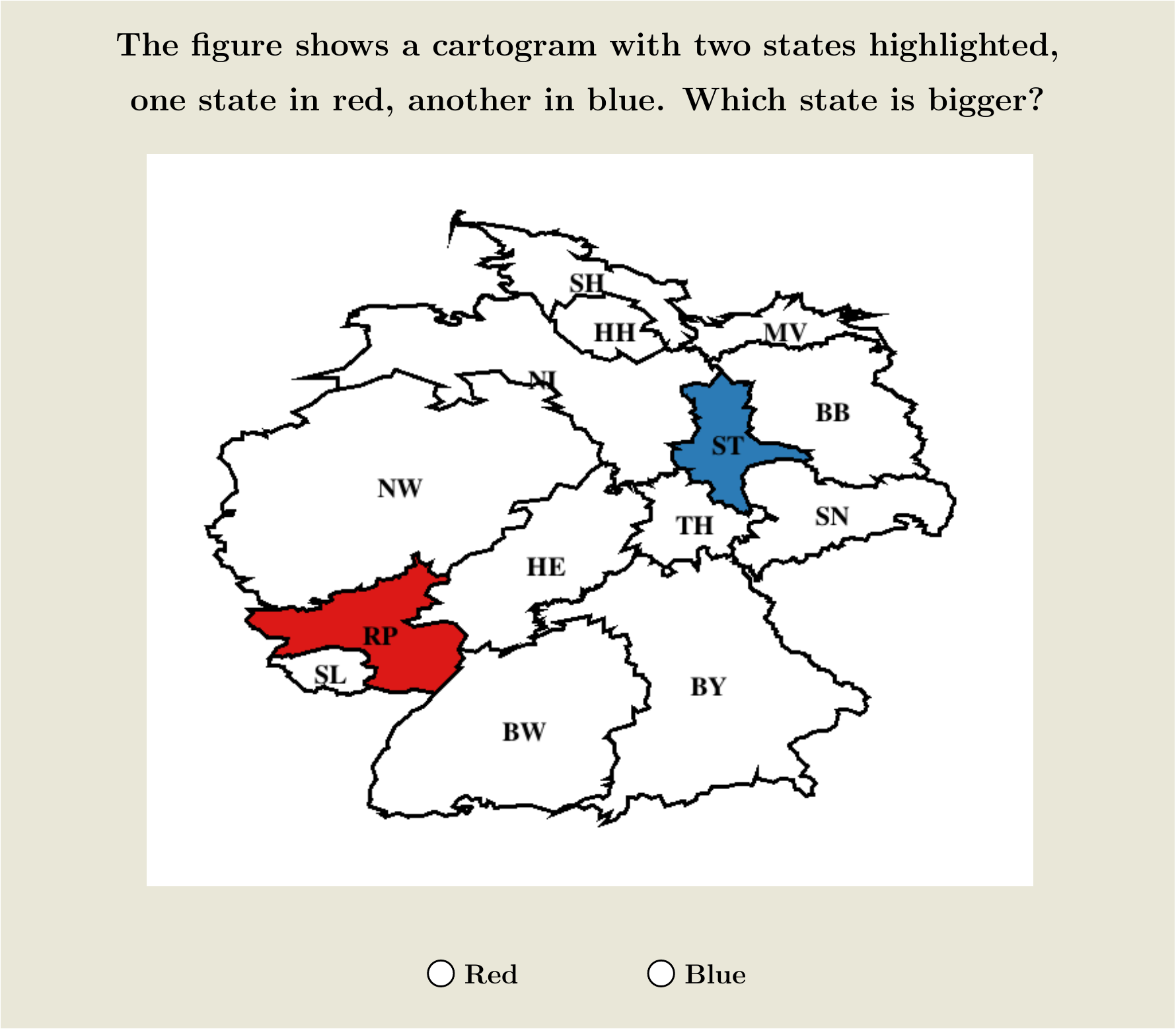}
\hfill
\includegraphics[width=0.48\textwidth]{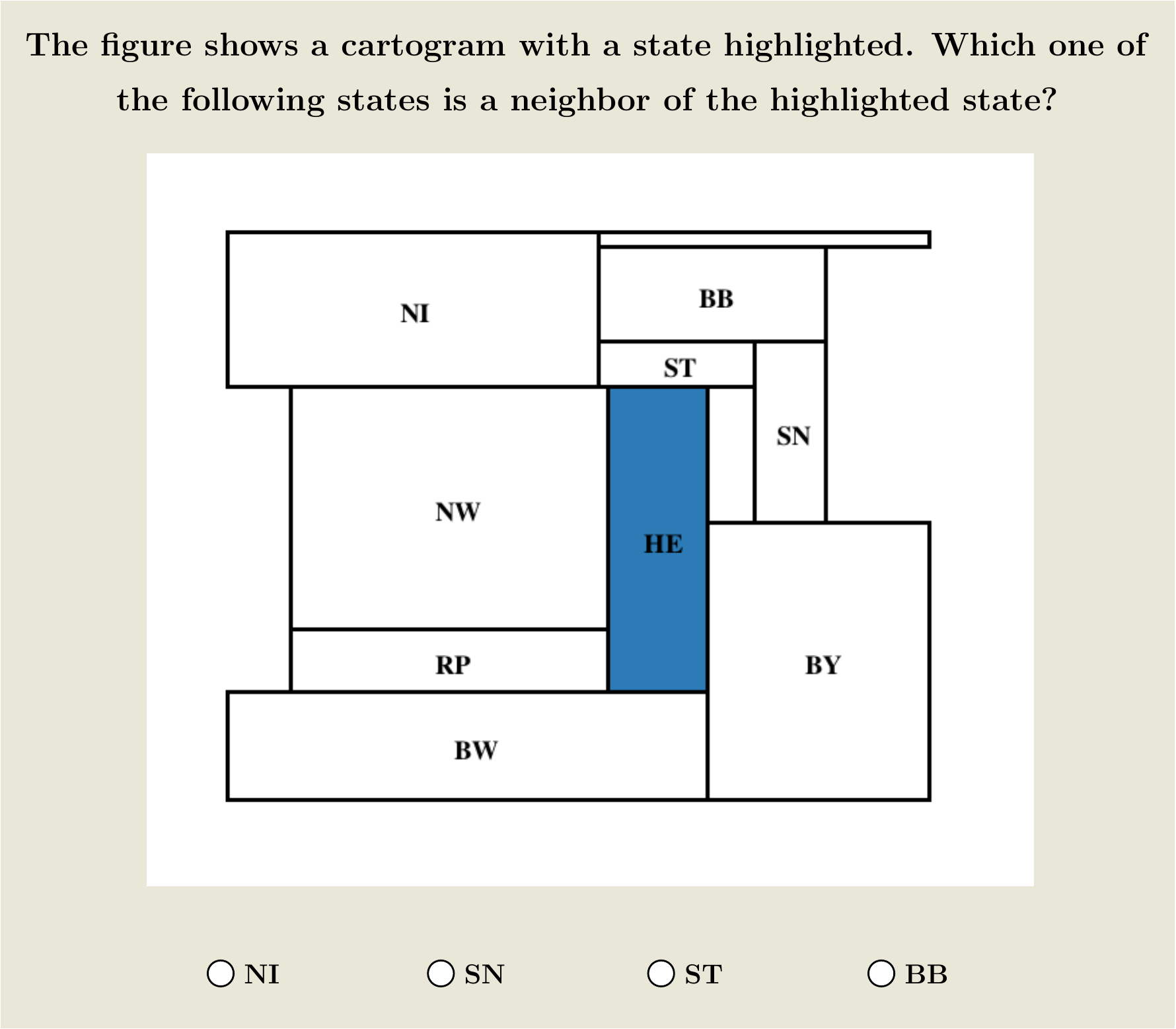}
\hfill
\\

(a) Contiguous cartogram, \textit{Compare} task \hspace{0.2\textwidth}(b) Rectangular cartogram, \textit{Find adjacency} task\\

\medskip

\hfill
\includegraphics[width=0.48\textwidth]{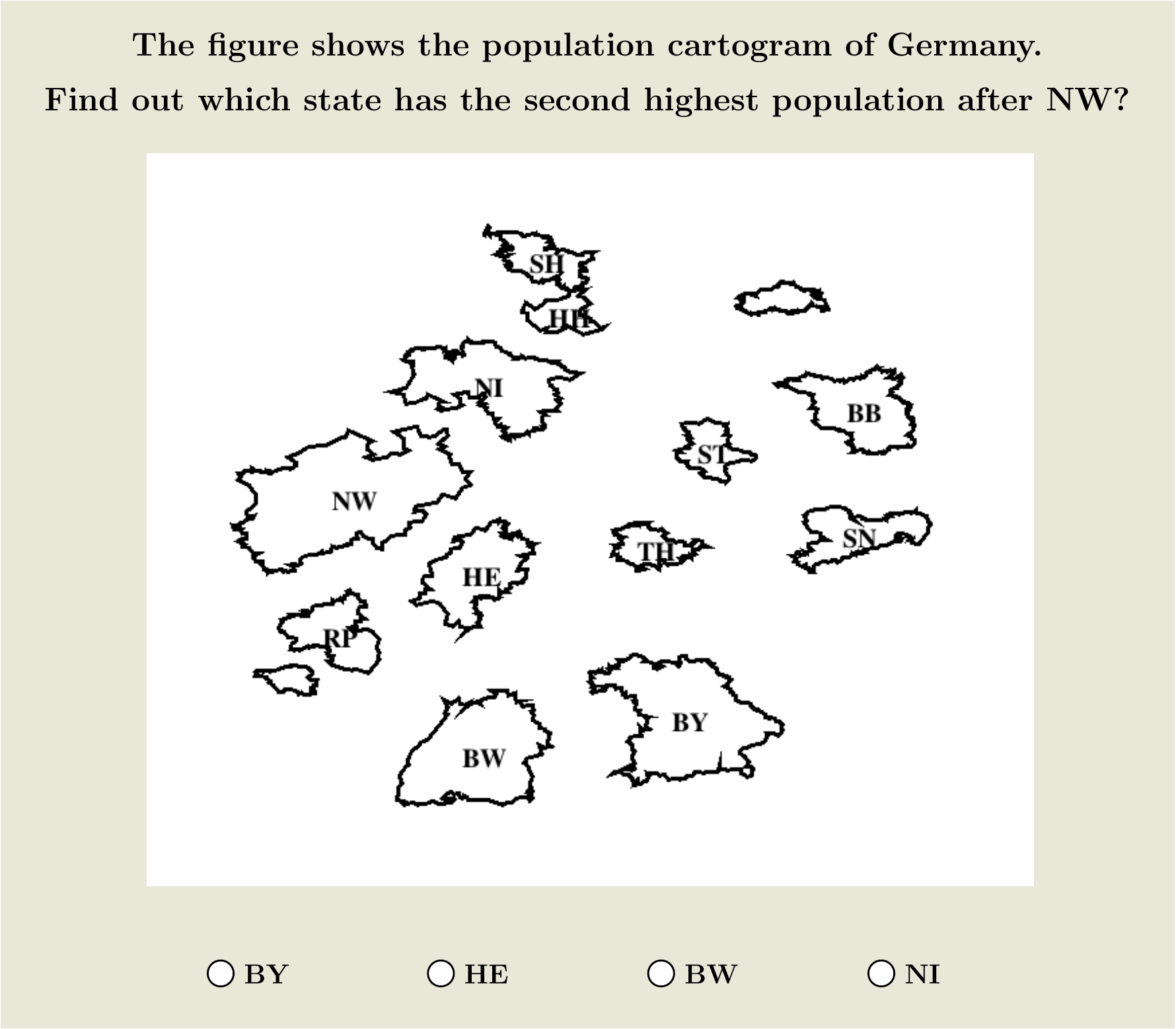}
\hfill
\includegraphics[width=0.48\textwidth]{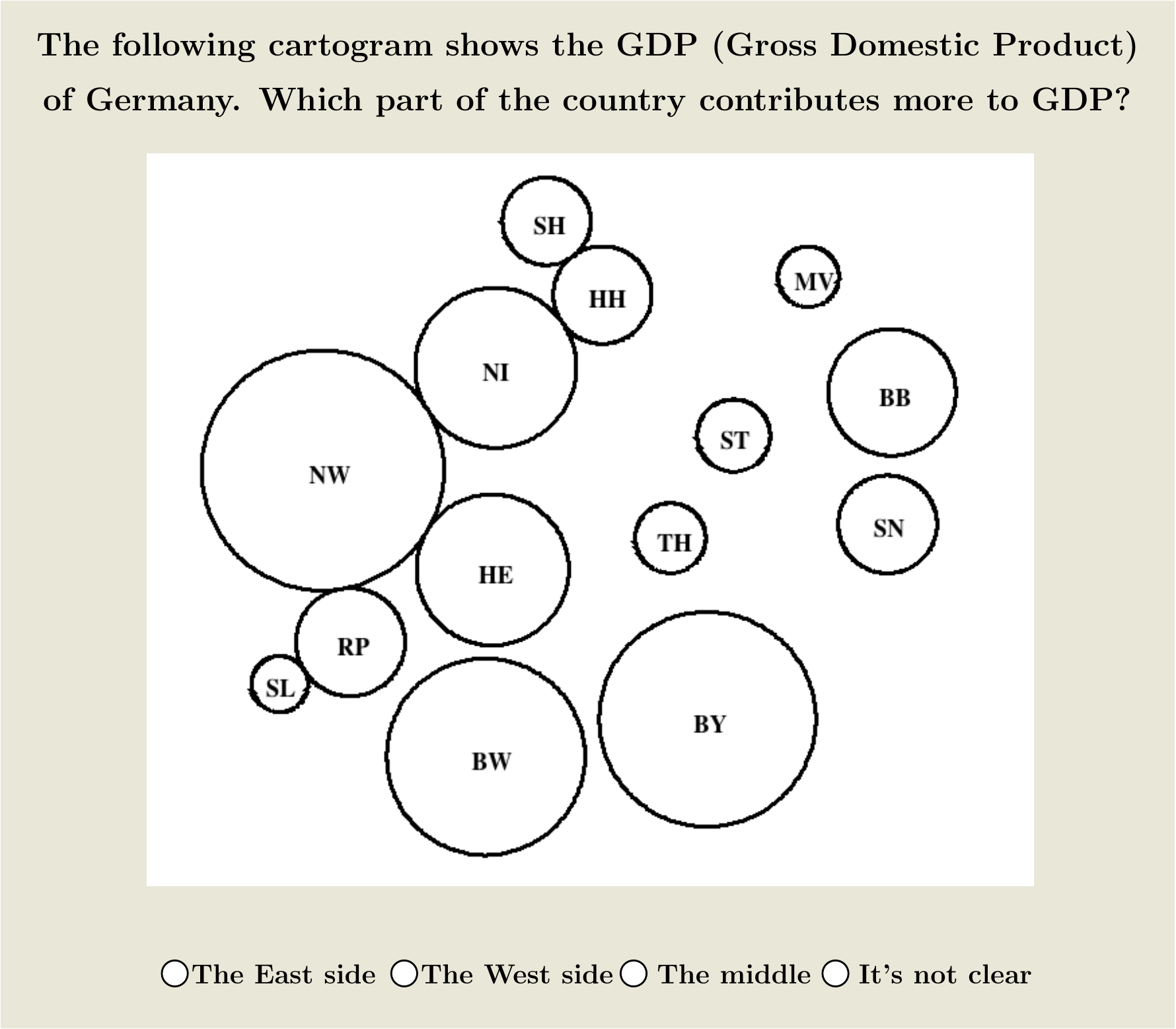}
\hfill
\\

\hspace{-0.08\textwidth}
(c) Non-contiguous cartogram, \textit{Find top-$k$} task \hspace{0.15\textwidth}(d) Dorling cartogram, \textit{Summarize} task\\

\caption{Example tasks on four types of cartograms of Germany.}

\label{fig:four-types}
\end{figure*}


Graphical perception of area is  relevant to cartograms as different methods generate different shapes (circles, rectangles, irregular polygons). 
There is a great deal of research in visualization and cartography about the impact of length, area, color, hue, and texture on map visualization and understanding.
Bertin~\cite{BERTIN83} was one of the first to provide systematic guidelines to test visual encodings.
Cleveland and McGill~\cite{cleveland1984graphical} extended Bertin's work with human-subjects experiments that established a significant accuracy advantage for position judgments over both length and angle judgments, which in turn proved to be better than area judgments. 
Stevens~\cite{Steven_law} modeled the mapping between the physical intensity of a stimulus and its perceived intensity as a power law. His experiments showed that subjects perceive length with minimal bias, but underestimate differences in area. This finding is further supported by Cleveland et al.~\cite{cleveland1982judgments}. In a more recent study, Heer and Bostock~\cite{JM10} investigated the accuracy of area judgment between rectangles and circles, both of which provide similar judgment accuracy, but are worse than length judgments. 
These results were consistent with the findings about  ``judgment of size'' by Teghtsoonian~\cite{teghtsoonian1965judgment}. Dent~\cite{dent1975} surveyed related work in magnitude estimation and suggested that the shapes of the enumeration units in cartograms should be irregular polygons or squares. 
However, it is difficult to use these experiments directly to determine what would work best in the cartogram setting, as the datasets used, the tasks given, and the experimental conditions vary widely from experiment to experiment.

\section{Cartogram Types}
\label{sec:types}

There is a wide variety of algorithms that generate cartograms and three major design dimensions along which cartograms vary: 
\begin{itemize}
\item \textbf{Statistical accuracy:} how well do the modified areas represent the corresponding statistic shown (e.g., population or GDP). This is measured in terms of ``cartographic error.''

\item \textbf{Geographical accuracy:} how much do the modified shapes resemble the original geographic shapes and how well preserved are their relative positions.

\item \textbf{Topological accuracy:} how well does the topology (as measured by adjacent regions) of the cartogram match that of the original map.
\end{itemize}

There is no ``perfect'' cartogram that is geographically accurate, preserves the topology, and also has zero cartographic error~\cite{AKV15}. Some cartograms preserve shape at the expense of cartographic error, others preserve topology, still others preserve shapes and relative positions. Cartograms can be broadly categorized in four types~\cite{ks07}: contiguous, non-contiguous, Dorling, and rectangular; see Fig.~\ref{fig:four-types}.

\textbf{Contiguous Cartograms: }
These cartograms deform the regions of a map, so that the desired areas are obtained, while
 adjacencies are maintained; see Fig.~\ref{fig:four-types}(a). 
They are also called \textit{deformation cartograms}~\cite{AKV15}, since
 the original geographic map is modified (by pulling, pushing, and stretching the boundaries) to change
 the areas of the regions on the map. Worldmapper~\cite{WorldMapper} has a rich collection of diffusion-based cartograms.
Among deformation cartograms the most popular variant is the ones generated by the diffusion-based algorithms of Gastner and Newman~\cite{GN04}, which we use in our evaluation. Others of this type include the rubber-map cartograms by Tobler~\cite{Tobler73}, contiguous area cartograms by Dougenik et al.~\cite{DCN85}, 
CartoDraw by Keim et al. ~\cite{KNPS03}, constraint-based continuous area cartograms by House and Kocmoud~\cite{HK98}, and medial-axis-based cartograms by Keim et al. ~\cite{KPN05}.

In deformation cartograms, since the input map is deformed to realize some given weights, the original map is often  recognizable, but the shapes of some countries might be distorted.
Recent variants for contiguous cartograms allow for some cartographic error in order to better preserve shape and topology~\cite{zackary_blog}.

\textbf{Rectangular Cartograms:} 
Rectangular cartograms schematize the regions in the map with rectangles; see Fig.~\ref{fig:four-types}(b). 
These are ``topological cartograms'' where the topology of the map (which country is a neighbor of which other country) is represented by the dual graph of the map, and that graph is used to obtain a schematized representation with rectangles. In rectangular cartograms there is often a trade-off between achieving zero (or small) cartographic error and preserving the map properties (relative position of the regions, adjacencies between them). 

Rectangular cartograms have been used for more than 80 years~\cite{Raisz34}. Several more recent methods for computing rectangular cartogram have also been proposed~\cite{BSV12, hkps04, ks07}. In our study, we use a state-of-the-art rectangular cartograms algorithm~\cite{BSV12}. There are several options for this type of algorithm and we choose the variant where the generated cartogram preserves topology (adjacencies), at the possible expense of some cartographic error.
Note that in addition to possible cartographic errors in this particular variant,  rectangular cartograms in general have one major problem. To make a map realizable with a rectangular cartogram, it might be necessary to merge two countries into one (which is highly undesirable in practice), or  
to split one country into two parts~\cite{ks07}. When recombining them this leads to regions that are no longer rectangular. In our study, we used the variant where the regions remain rectangular, at the expense of some countries getting merged with other countries. In particular 5 states in the map of USA, 3 states in Germany and 2 regions in Italy get merged in this algorithm. While some countries have states and others have provinces and regions, for simplicity we refer to all of them as ``regions'' in the rest of the paper.

\textbf{Non-Contiguous Cartograms:} 
These cartograms are created by starting with the regions of a map, and scaling down each region independently, so that the desired size/area is obtained; see Fig.~\ref{fig:four-types}(c). 
They satisfy area and shape constraints, but do not preserve the topology of the original map~\cite{Krauss_ms}. 
The non-contiguous cartograms method of Olson~\cite{Olson} scales down each region in place (centered around the original geographical centroid), while preserving the original shapes. For each region, the density (statistical data value divided by geographic area) is computed and the region with the highest density is chosen as the anchor, i.e., its area remains unchanged while all other regions become smaller in proportion to the given statistical values. If the highest density region is geographically small, there will be a lot white space in the cartogram. If this is the case, Olson's method searches for a high density region of reasonable size as an anchor; in this case smaller regions with higher densities are enlarged rather than reduced. In our study, we optimize the choice for an anchor to ensure that no pairs of regions overlap.
Despite these efforts to reduce white space, since the size of the final regions depends on their original size and statistic to be shown, some regions may become too small. 
By definition, non-contiguous cartograms do not preserve the original region adjacencies, however, there is some evidence that the loss of  adjacencies might not cause serious perceptual difficulties~\cite{KPN05}.

\textbf{Dorling Cartograms:} 
Dorling cartograms represent areas by circles~\cite{dorling96}.
Data values are realized by size of the circle: the bigger the circle, the larger the data value; see Fig.~\ref{fig:four-types}(d). 
 However, in order to avoid overlaps, circles might need to be moved (typically as little as possible) away from their original geographic locations. Unlike contiguous and non-contiguous cartograms, Dorling cartograms preserve neither shape nor topology.
Dorling cartograms became very popular in the UK where the computer programs for generating Dorling cartograms were first published by its creator Danny Dorling. 
Dorling-style cartograms have become very popular on the web with JavaScript D3 implementations.

\section{Metric-Based Analysis}

We performed a comparative study on the four major types of cartograms, based on a set of quantitative performance metrics.
Various quantitative cartogram measures have been proposed in the literature, and several studies used ad-hoc definitions of performance metrics to compare new algorithms to existing ones~\cite{BSV12,KNPS03,ks07,BMS10}.  A recent standard set of such parameters with which to compare and evaluate cartograms~\cite{AKV15}, can be categorized based on the three cartogram dimensions:

\begin{figure}[h]
\centering
\parbox{0.5\textwidth}
{\hspace{-0.4cm}
\includegraphics[width=0.25\textwidth]{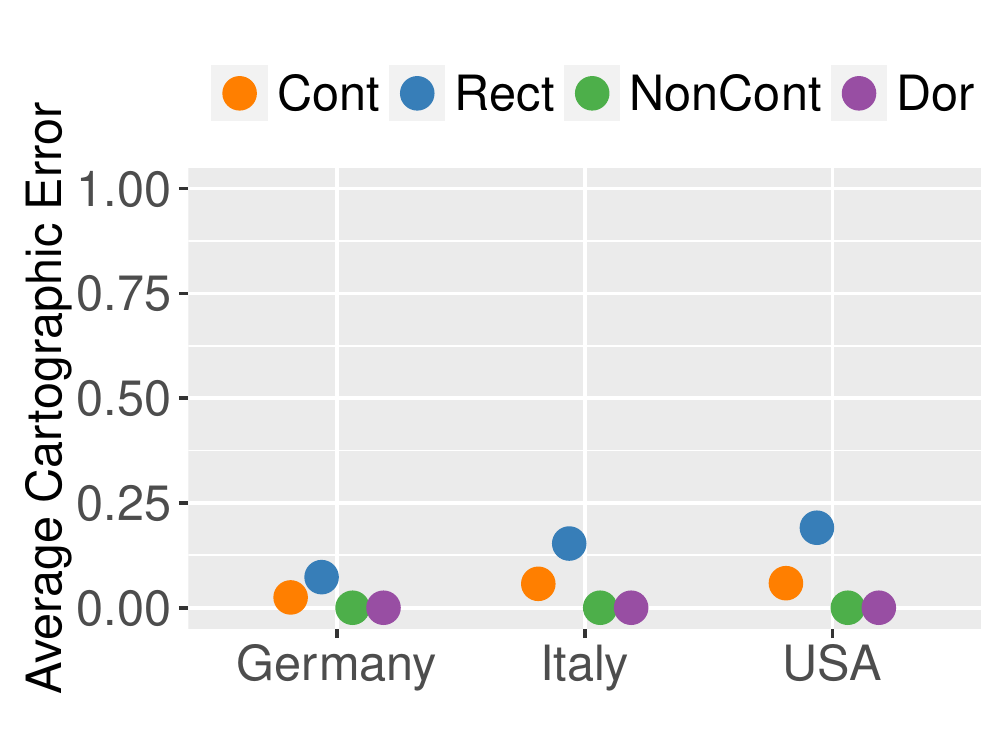}
\includegraphics[width=0.25\textwidth]{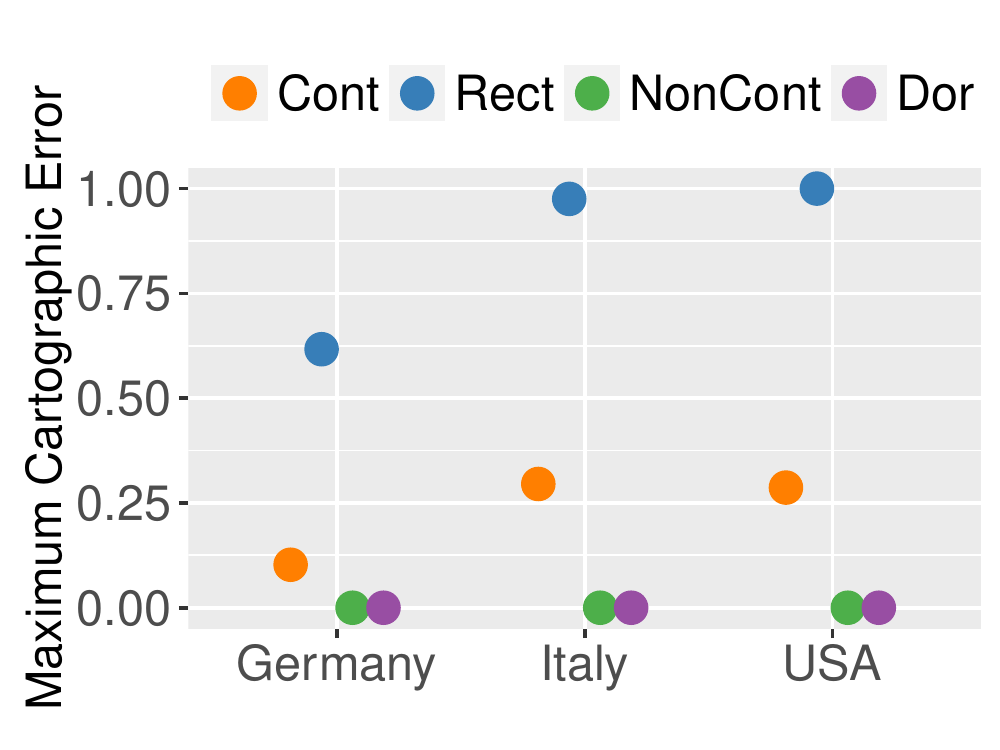}
}\\
(a) Statistical accuracy metrics\\
\parbox{0.5\textwidth}
{\hspace{-0.4cm}
\includegraphics[width=0.25\textwidth]{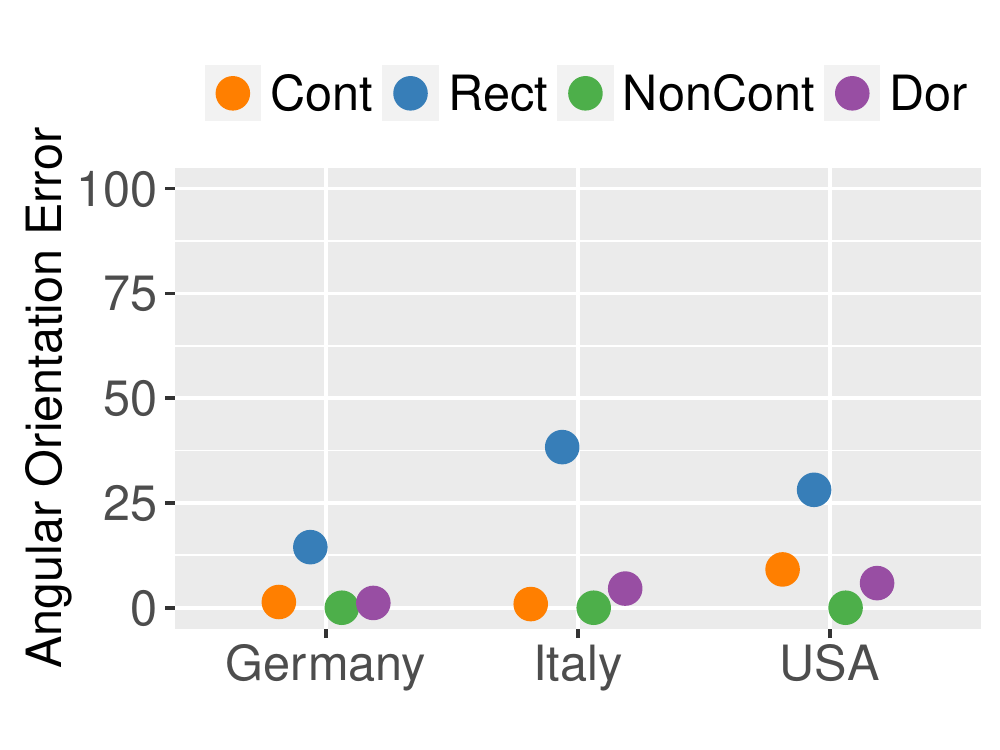}
\includegraphics[width=0.25\textwidth]{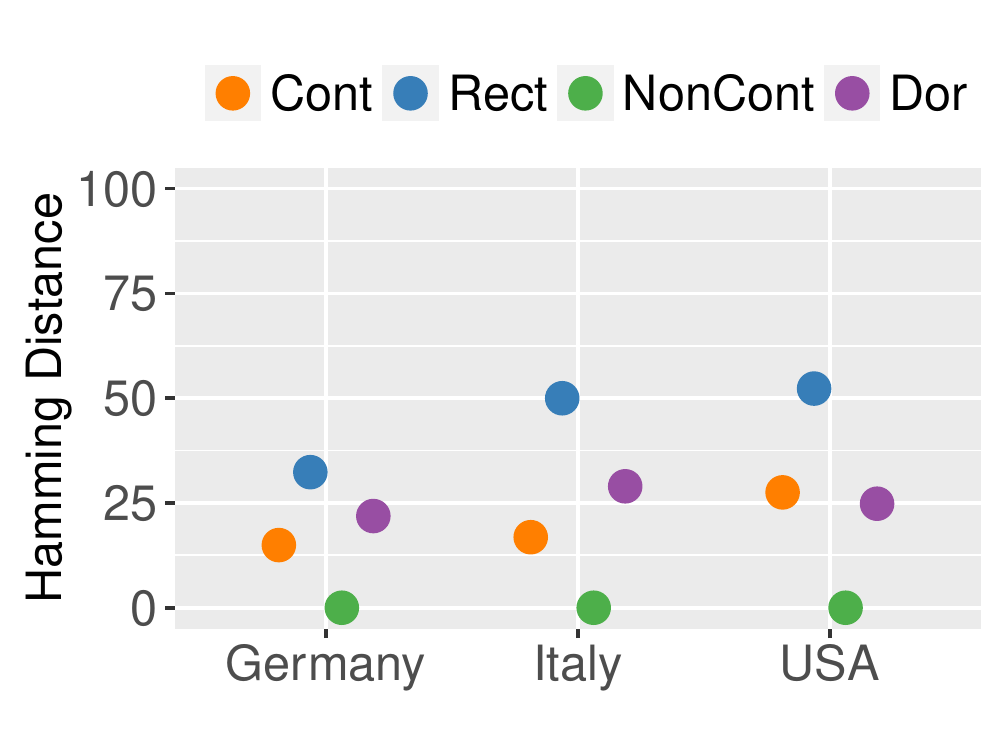}
}\\
(b) Geographical accuracy metrics\\
\caption{\small Metric-based comparison of four cartogram types, using cartograms of Germany, Italy and USA. (a) Metrics for statistical accuracy: average cartographic error (left) and maximum cartographic error (right), (b) metrics for geographical accuracy: angular orientation error (left) and Hamming distance (right).}
\label{fig:metric}
\end{figure}

\textbf{Statistical Accuracy:}
This measures how well the obtained region areas in the cartogram match the desired statistical values. 
The \textit{cartographic error} for a region $v$ in the cartogram is defined as $\frac{|o(v)-w(v)|}{max\{o(v),w(v)\}}$, where $o(v)$ and $w(v)$ are the obtained and desired area for the region. After evaluating different options for measuring the cartographic error of a given cartogram~\cite{KNPS03,KNP04,BSV12}, Alam et al.~\cite{AKV15} advocate for both the average error and the maximum error, as measures of statistical distortion in the cartograms.

\label{stat}
\textbf{Geographical Accuracy: } 
Two measures are also proposed in this context: one for region shape preservation and another for the preservation of the relative positions of the regions. 
Shape preservation is measured using the Hamming distance~\cite{SKI98}, also known as the symmetric difference~\cite{MRS10} between two polygons. The polygons for each cartogram region and the corresponding map region are normalized to unit area and superimposed on top of each other; the fraction of the area in exactly one of the polygons is the Hamming distance $\tsdist$. Relative position preservation is measured by the \textit{angular orientation error}, $\angular$, defined by Heilmann \textit{et al.}~\cite{hkps04} and obtained by computing the average change in the slope of the line between the centroids of pairs of regions.

\textbf{Topological Accuracy:} 
Topological accuracy is measured with the \textit{adjacency error} $\toperror$: the fraction of the regional adjacencies that the cartogram fails to preserve, i.e., $\toperror = 1 - \frac{|E_c\cap E_m|}{|E_c\cup E_m|}$, where $E_c$ and $E_m$ are respectively the adjacencies between regions in the cartogram and the original map.
\medskip

Alam et al.~\cite{AKV15} used these measures to compare five cartogram algorithms.
Among these five were contiguous and rectangular cartograms, but not Dorling and non-contiguous cartograms. We add these two cartogram types and evaluate their performance with three different countries (Germany, Italy, USA) and with two different statistics (population and GDP) for each map. Fig.~\ref{fig:metric} shows the results for statistical and geographical accuracy, for each of the three countries.

Statistical Accuracy: Dorling and non-contiguous cartograms are perfect in that regard, while rectangular cartogram have 3--10 times greater cartographic error than diffusion cartograms; see Fig.~\ref{fig:metric}(a).

Geographical Accuracy: Non-contiguous cartograms are perfect in that regard (zero angular orientation error and Hamming distance), while contiguous cartograms show low errors in both shapes and angles. Rectangular cartograms are a clear outlier with errors in both shapes and angles that are at least 2 times greater than any other cartogram type; see Fig.~\ref{fig:metric}(b).

Topological Accuracy: Contiguous cartograms are perfect, and so are the topology-preserving variant of rectangular cartograms. Non-contiguous cartograms do not maintain any adjacencies. Dorling cartograms have high adjacency error, especially the variant with attraction forces keeping the regions near the correct geographic locations. 
We note that adjacency error might not be a ``fair" metric for non-contiguous and Dorling cartograms, since for both of these two cartogram types, the region becomes non-contiguous and geographical proximity rather than exact adjacency becomes a guide for topological relation.

We discuss these results, together with the results of the task-based study, in Section~\ref{sec:results}.

\section{Visualization Tasks in Cartograms}
Cartograms are employed to simultaneously convey two types of information: geographical and statistical.
 Our goal is to evaluate different types of cartograms in these two aspects, by conducting experiments that cover the spectrum of possible tasks. In this context, a recent task taxonomy for cartograms is particularly useful, as it categorizes tasks in different dimensions (e.g., goals, means, characteristics) and groups similar tasks together~\cite{Task_C}. 
 
In order to cover the spectrum of tasks, and yet to keep the number of tasks low for practical reasons, we selected seven of the ten tasks in the taxonomy. We included basic map tasks, such as \textit{find adjacency} and 
 \textit{recognize}. We also included basic statistical tasks, such as \textit{compare}, as well as composite tasks, such as \textit{summarize}. 
 
The tasks \textit{filter}, \textit{cluster}, and \textit{find top-$k$} are tasks that have similar goals (exploring data), similar means (finding data relation), similar high-level data characteristics, and all three tasks consider ``all instances'' of the data. 
Another group of tasks with similar goals and means contains \textit{summarize} and \textit{identify} tasks.
We used \textit{find top-$k$} and \textit{summarize} as representatives from these two groups of tasks.

Here we describe all the visualization tasks used in our study; these are also included in Table~\ref{tab:tasks}, where  the exact input setting, along with the exact questions given to the participants, are summarized.

\textbf{Compare:}
The \textit{compare} task has been frequently used in taxonomies and evaluations~\cite{Jen, RR13, Weh93}. The task typically asks for similarities or differences between attributes; see 
Fig.~\ref{fig:four-types}(a) for an example of a \textit{compare} task in our experiment.

\textbf{Detect change:} 
In cartograms the size of a region is changed in order to realize the input weights. Since change in size (i.e., whether a region has grown or shrunk) is a central feature, it is crucial that the viewer be able to detect such change.

\textbf{Locate:}
The task in this context corresponds to searching and finding the position of a region in a cartogram. In some taxonomies this task is denoted as \textit{locate} and in others as  \textit{lookup}, but these are not necessarily the same~\cite{BM13}.
Since cartograms often drastically deform an existing map, even if the viewer is familiar with the underlying maps, finding something in the cartogram might not be a simple lookup. 

\textbf{Recognize:}
One of the goals in generating cartograms is to keep the original map recognizable, while distorting it to realize the given statistic. Therefore, this is an important task in our taxonomy. The aim of this task is to find out if the viewer can recognize the shape of a region from the original map when looking at the cartogram. 

\textbf{Find top-$k$:}
This is another commonly used task in visualization. Here the goal is to find $k$ entries with the maximum
 (or minimum) values of a given attribute. This task generalizes tasks,
 such as \textit{Find extremum} and 
 \textit{Sort}. In our evaluation, we ask the subjects to find out the region with the highest or second highest value of an attribute; see Fig.~\ref{fig:four-types}(c).

\textbf{Find adjacency:}
Some cartograms preserve topology, others do not. In order to understand the map characteristics properly, it is important to identify the neighboring regions of a given region.

\textbf{Summarize (Analyze / Compare Distributions and Patterns):}
Cartograms are most often used to convey the ``big picture''. \textit{Summarize} tasks ask the viewer to find patterns and trends in the cartogram.

\newcommand{\plotStudy}[2]
{
	\parbox{0.235\textwidth}
	{
		\smallskip
		\hspace{-0.2cm}
		\includegraphics[width=0.25\textwidth]{#1.pdf}
	}
}

\begin{table*}
\begin{tabular}{|c|c|c|c|c|c|c|}

\hline

& &
 \textbf{Input} & \textbf{Question} &
{\textbf{Time (s)}}
 & \textbf{Error \% }\\

\hline

{\rotatebox{90}{\textbf{H1}}}

& {\hspace{-0.15cm}\rotatebox{90}{{\hspace{-0.3cm}\textit{Locate}}}} &
\parbox{0.19\textwidth}{\small 
An original undistorted map is given and a state is highlighted in red. A cartogram of the map is shown.} 
 & \parbox{0.18\textwidth}{\small
 Locate this state in the cartogram.}
& \plotStudy{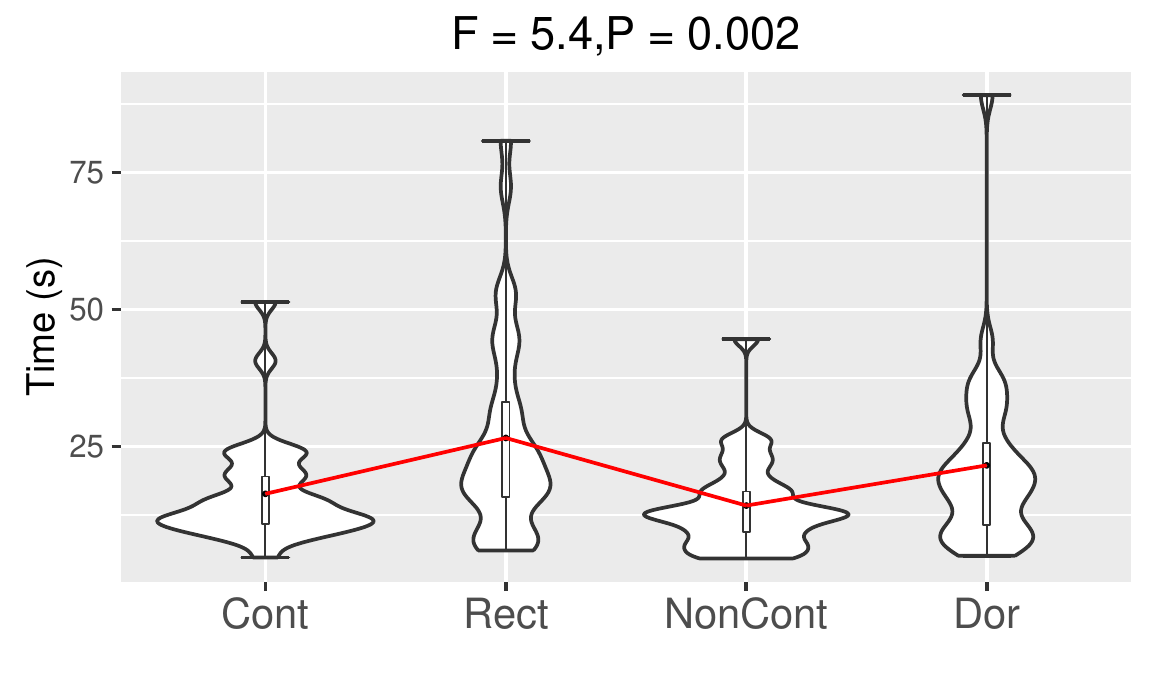}{0.74} & \plotStudy{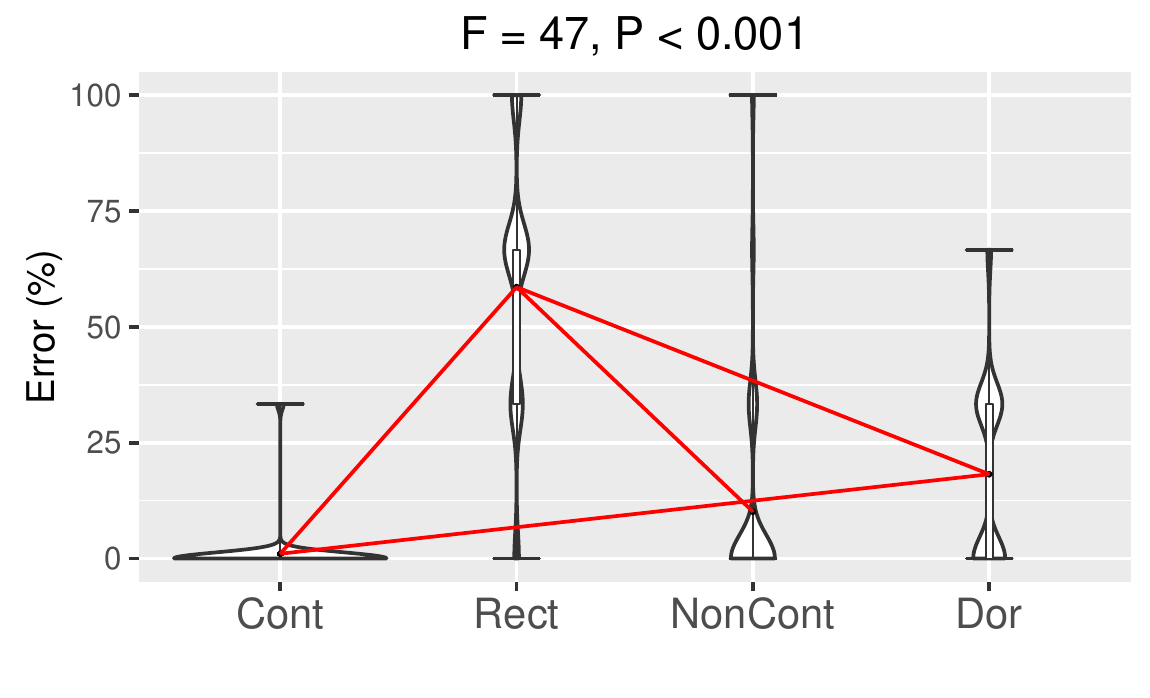}{0.74}\\

\hline
{\rotatebox{90}{\textbf{H2}}}
& {\hspace{-0.15cm}\rotatebox{90}{\hspace{-0.5cm}{\textit{Recognize}}}} &
\parbox{0.19\textwidth}{\small
 A state from the map of a country, and shapes of three states from the cartogram of that country are shown.} &

\parbox{0.18\textwidth}{\small
 Find out which cartogram state corresponds
	to the state from the original map.}
& \plotStudy{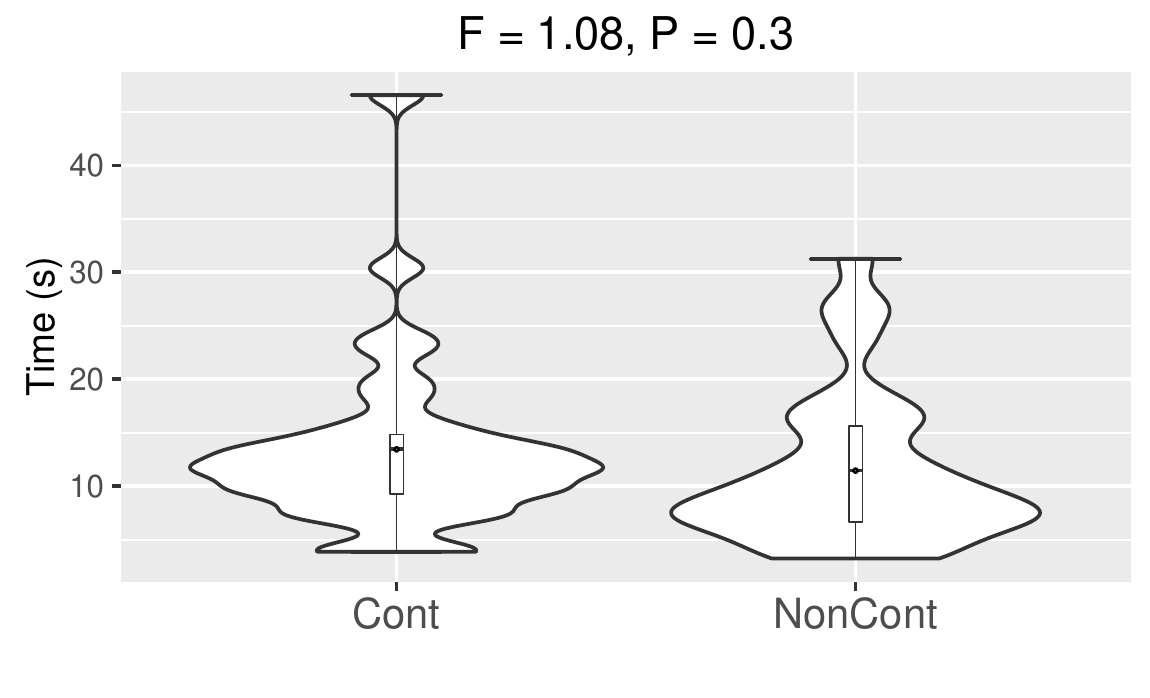}{0.74} & \plotStudy{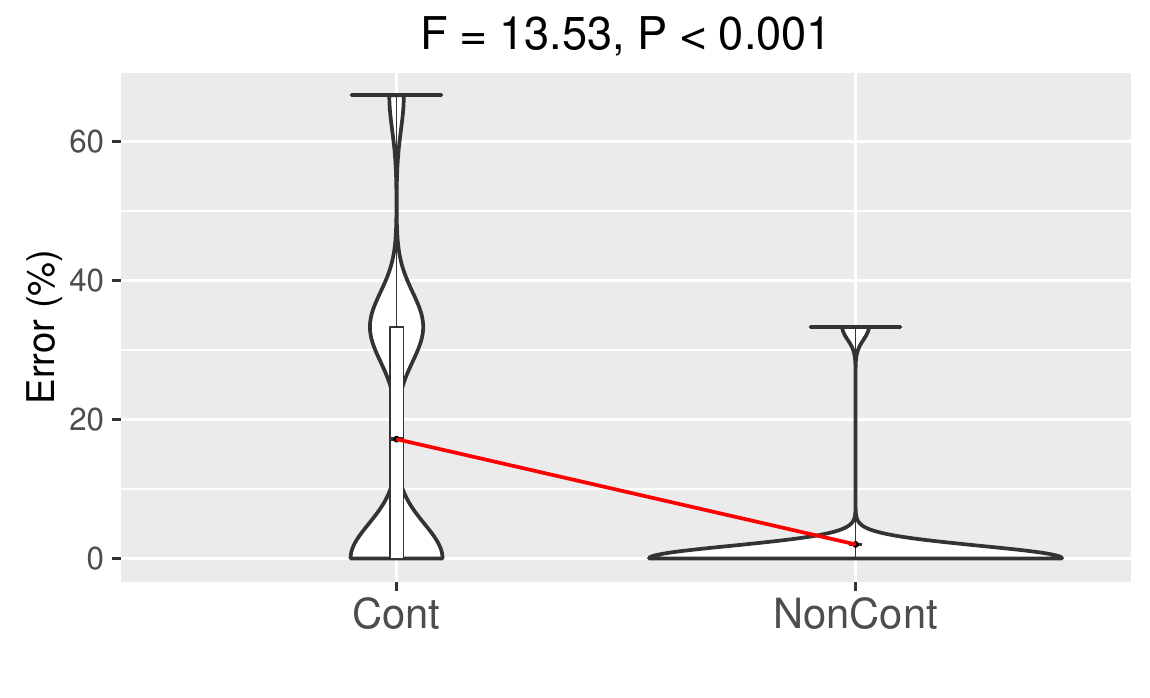}{0.74}\\

\hline


\multirow{3}{*}{\rotatebox{90}{\hspace{-2.5cm}\textbf{H3}}}

& {\hspace{-0.15cm}\rotatebox{90}{\hspace{-0.5cm}{\textit{Compare}}}} &
\parbox{0.19\textwidth}{\small
 A cartogram is shown with a red state and a blue state highlighted.}
 & \parbox{0.18\textwidth}{\small
 Which state is bigger: blue or red?}
& \plotStudy{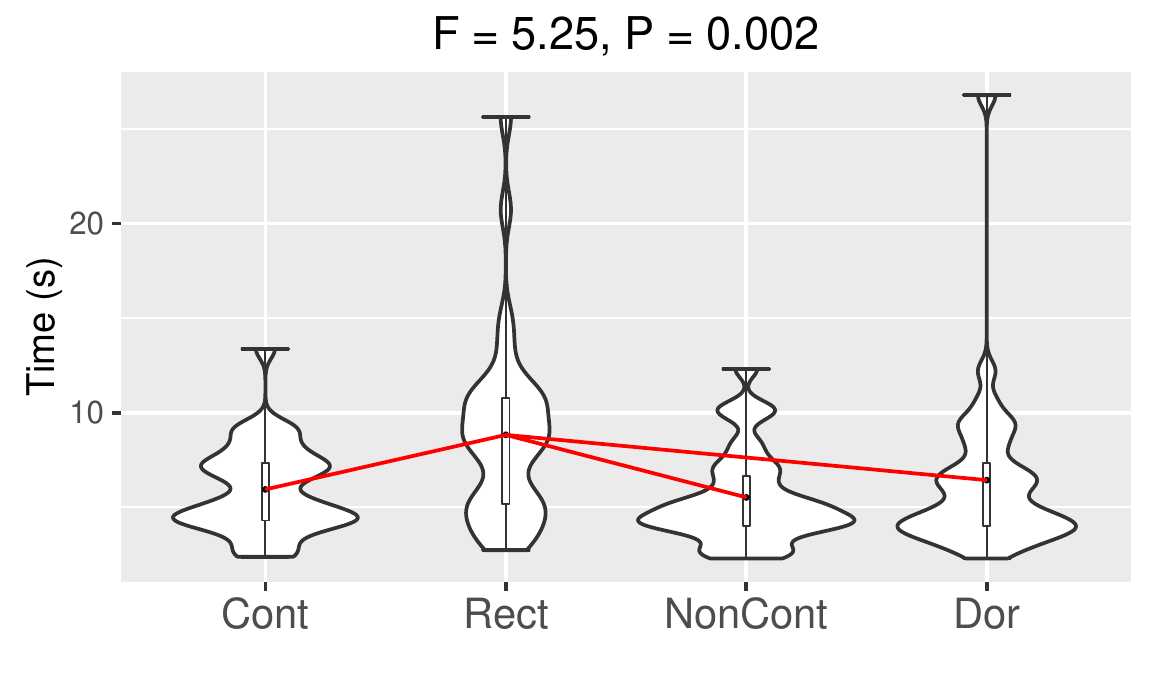}{0.74} & \plotStudy{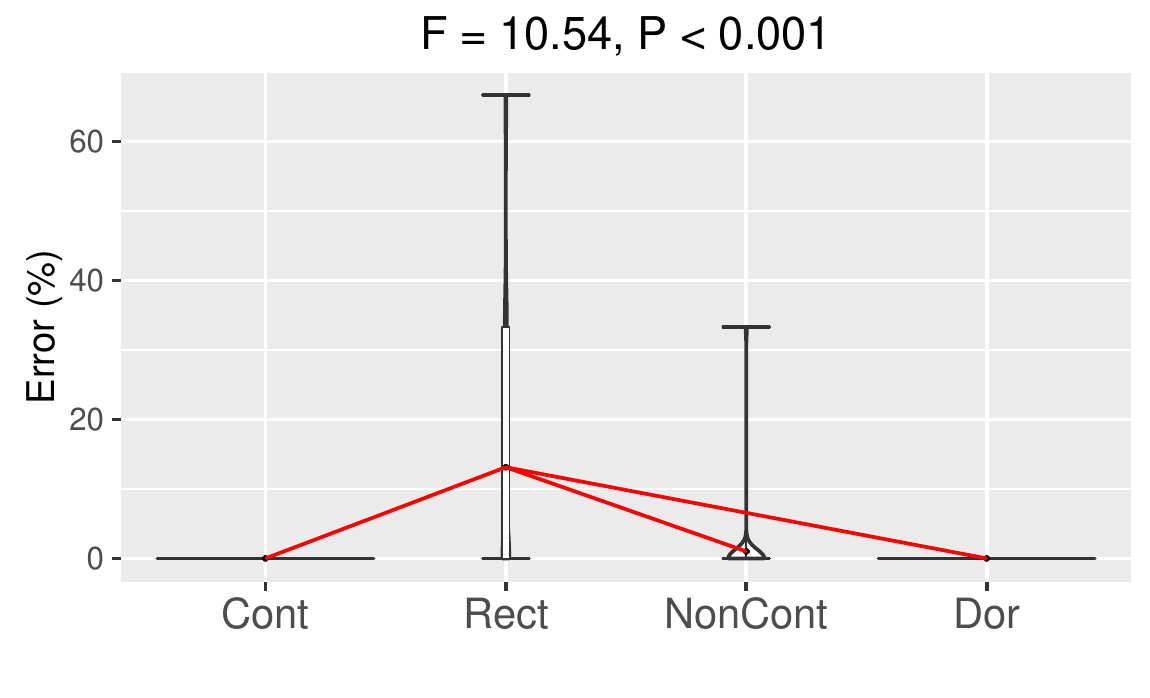}{0.74}\\

\cline{2-6}

 & {\hspace{-0.15cm}\rotatebox{90}{\hspace{-0.5cm}{\textit{Find top-$k$}}}} &
\parbox{0.19\textwidth}
{\small A cartogram of a country is shown.} &

\parbox{0.18\textwidth}{\small
 Find the state/region with the highest/second highest value of a statistic (e.g., population, GDP)}
& \plotStudy{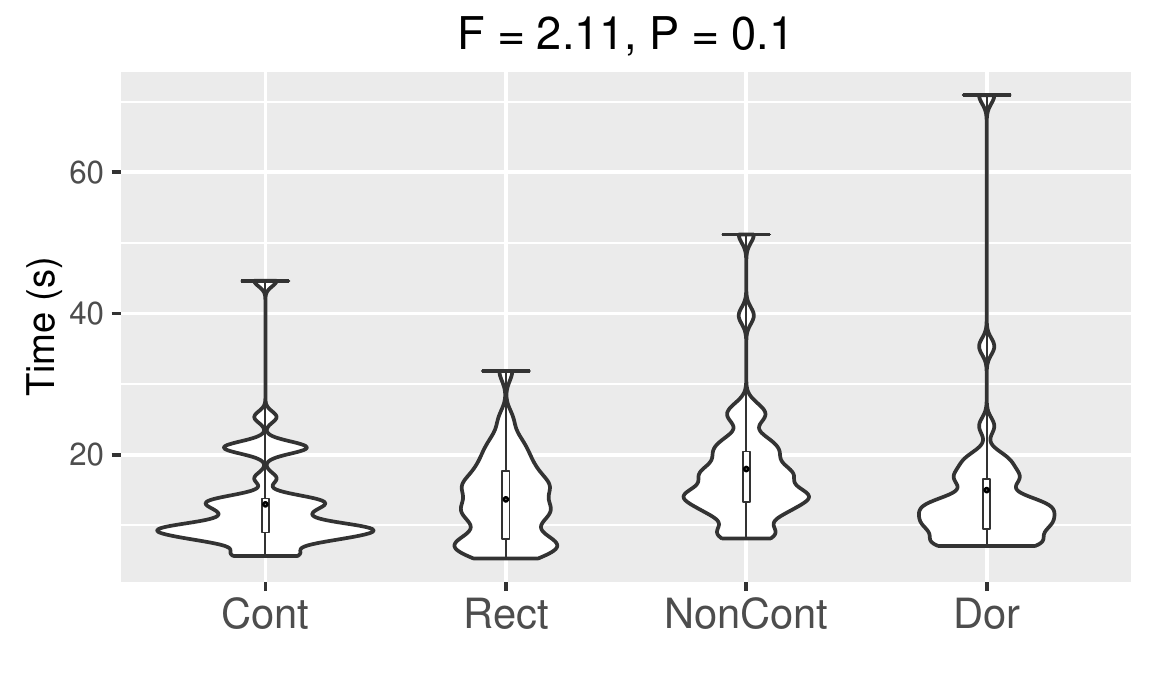}{0.74} & \plotStudy{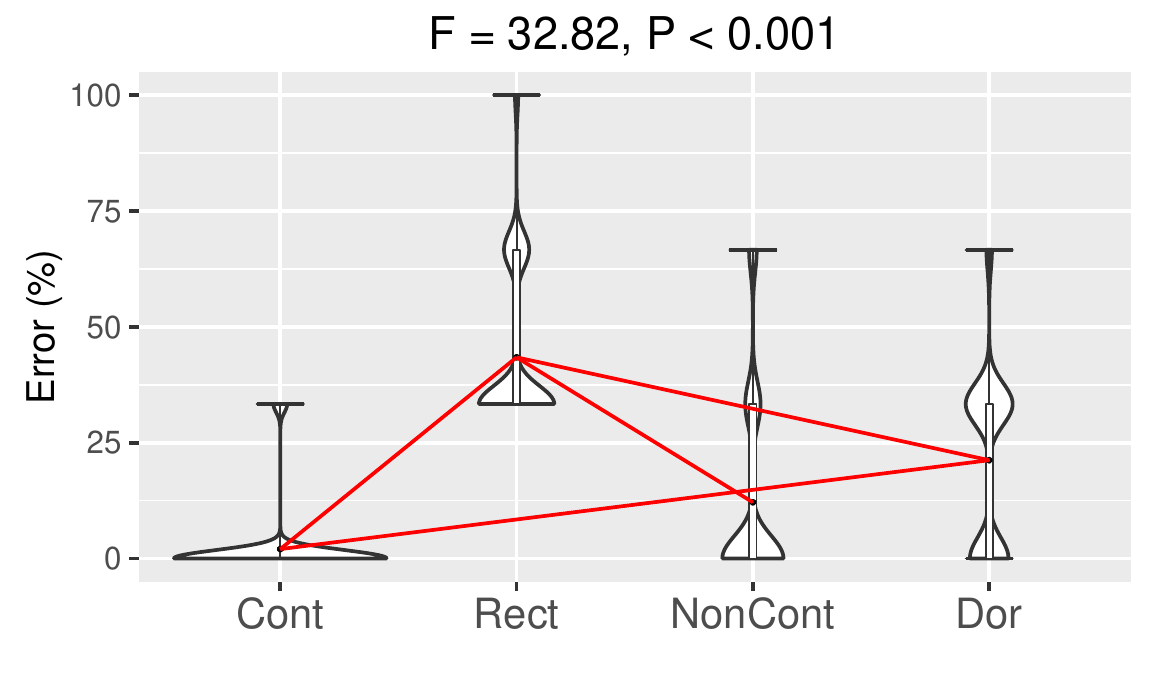}{0.74}\\

\cline{2-6}

& {\hspace{-0.15cm}\rotatebox{90}{\hspace{-0.7cm}{\textit{Detect Change}}}} &

\parbox{0.19\textwidth}{\small
 A map and a cartogram are shown. A state is highlighted in red on the map and in  blue on the cartogram.} &

\parbox{0.18\textwidth}{\small
 Compared to the red state in the map, has the blue state in the cartogram grown or shrunk?}
& \plotStudy{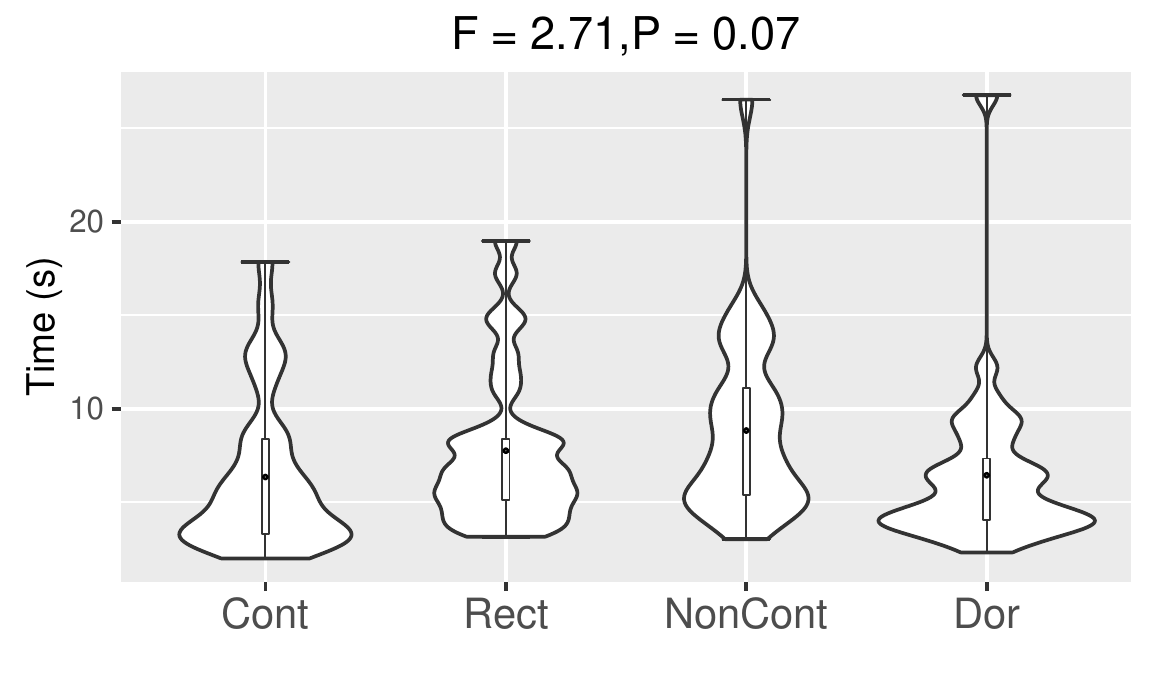}{0.74} & \plotStudy{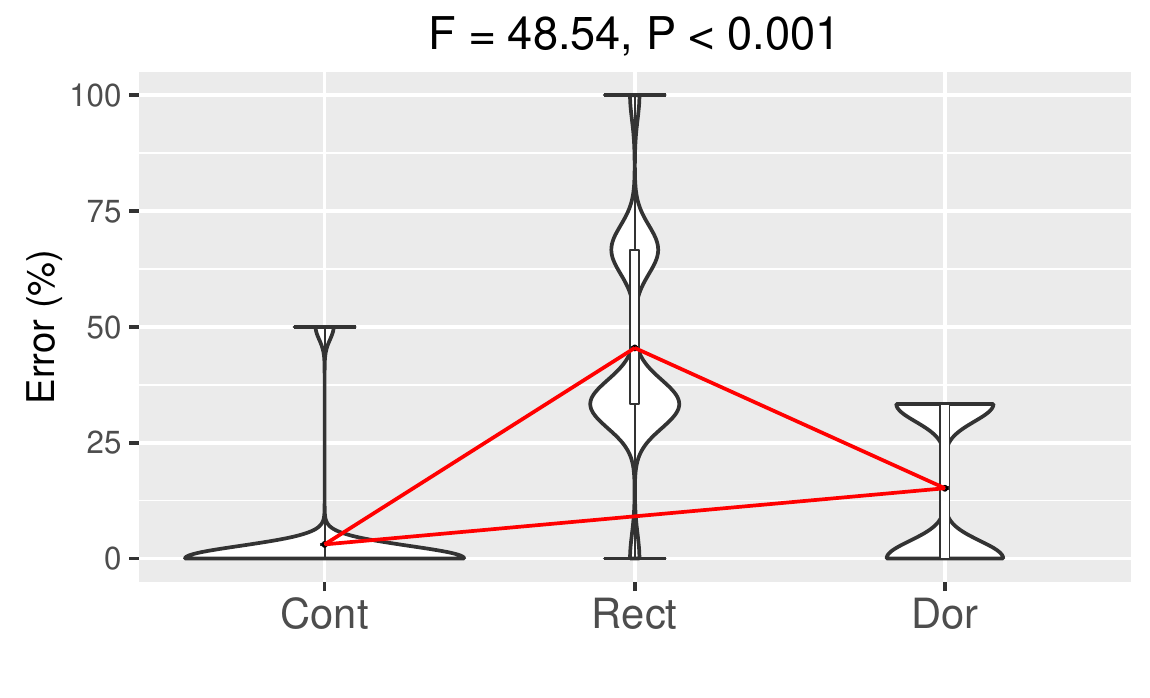}{0.74}\\

\hline

\rotatebox{90}{\textbf{H4}} & {\hspace{-0.15cm}\rotatebox{90}{\hspace{-0.8cm}{\textit{Find Adjacency}}}} &
\parbox{0.19\textwidth}{\small
 A cartogram is shown and a state is highlighted in red. A geographically undistorted map is given for reference.} 
 & \parbox{0.18\textwidth}{\small
 Which state is a neighbor of the highlighted state?}
& \plotStudy{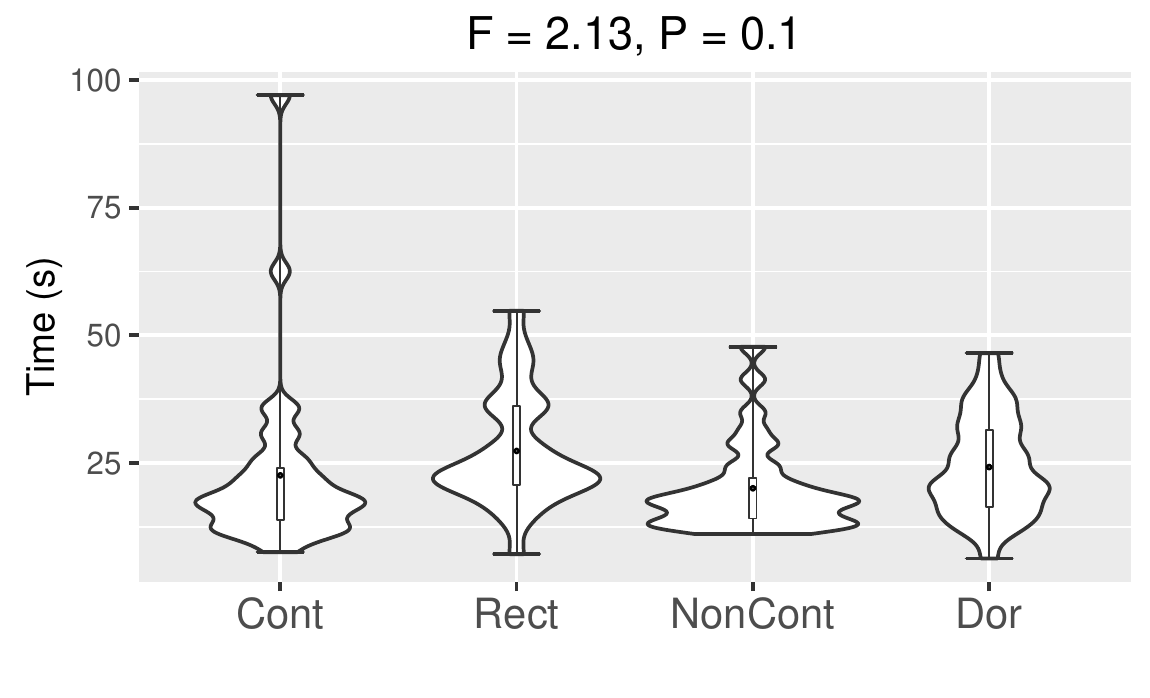}{0.74} & \plotStudy{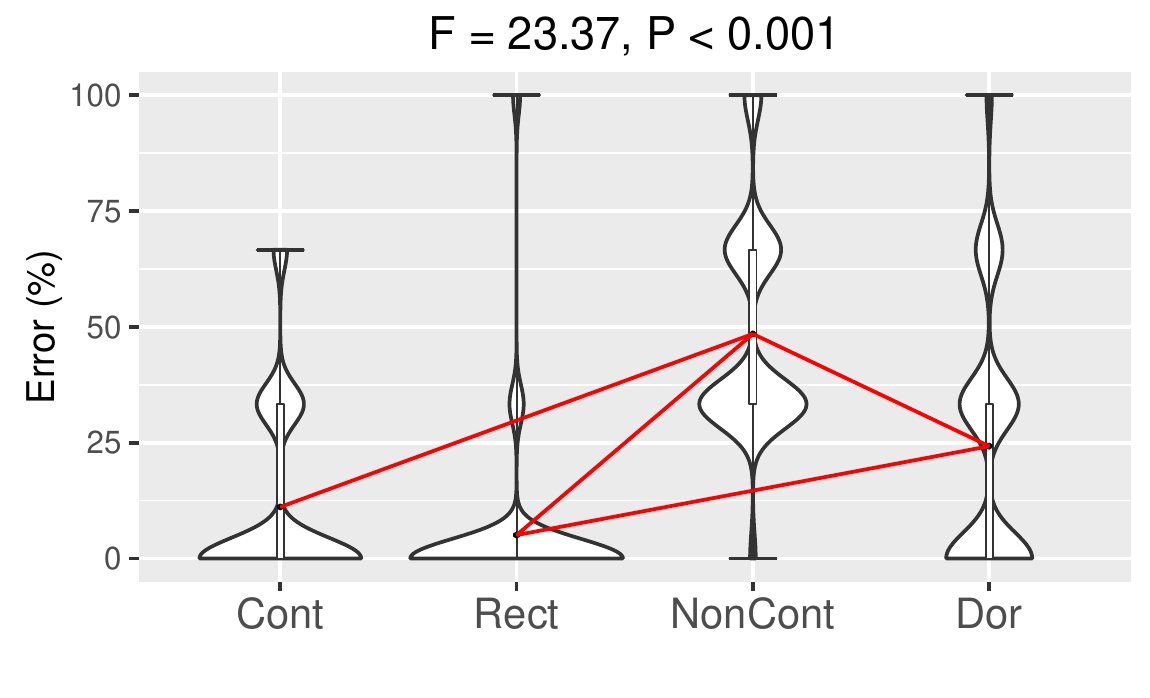}{0.74}\\

\hline

\multirow{4}{*}{\rotatebox{90}{\hspace{-1.2cm}\textbf{H5}}}
 & \multirow{4}{*}{{\hspace{-0.15cm}\rotatebox{90}{\hspace{-1.5cm}\textit{Summarize}}}} &
\parbox{0.19\textwidth}{\small
A cartogram of Italy shows the number of criminal incidents involving arson.
} &

\parbox{0.18\textwidth}{\small 
Where is this criminal activity high compared to other areas?
}
& \multirow{4}{*}{\plotStudy{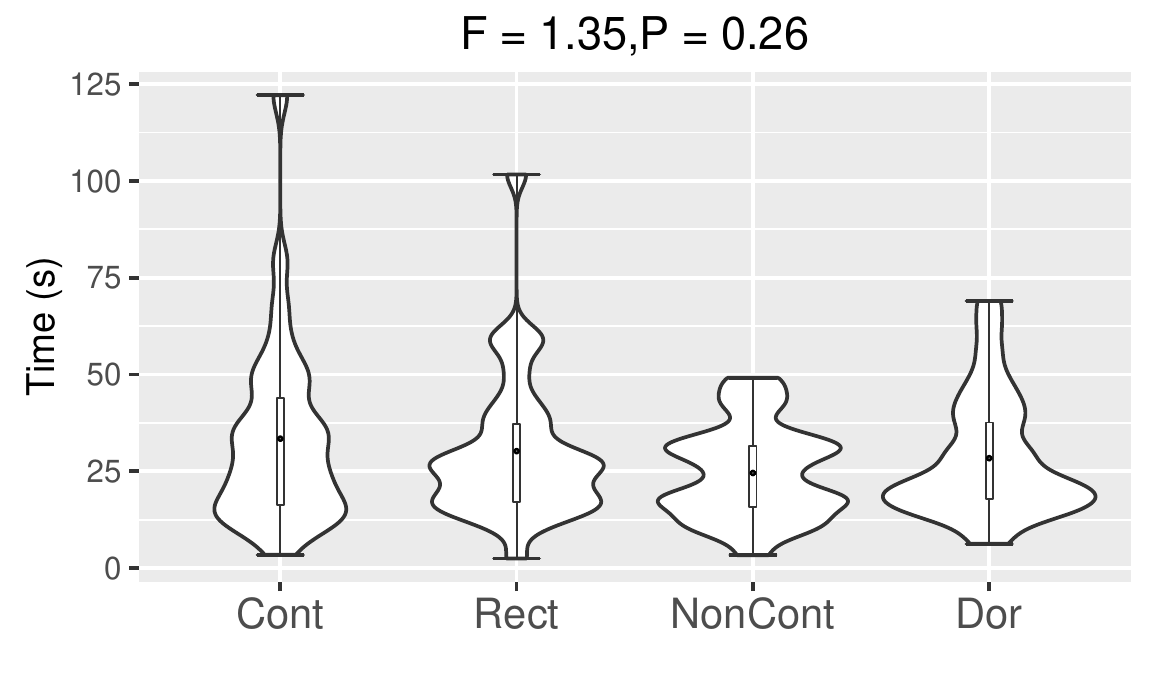}{0.74}} & \multirow{4}{*}{\plotStudy{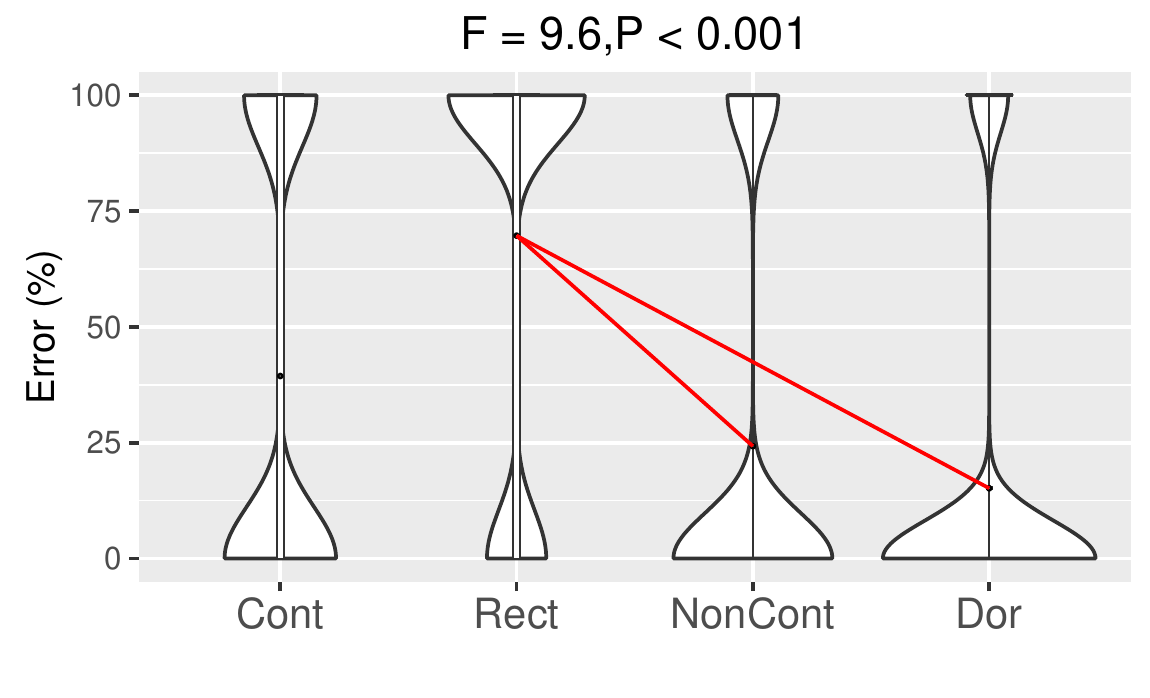}{0.74}}\\

\cline{3-4}

& & \parbox{0.19\textwidth}{\small
 A cartogram shows the GDP of Germany.}

&
\parbox{0.18\textwidth}
{
 \parbox{0.19\textwidth}{\small
 Which part of the country contributes more to GDP?}
}
 & & \\

\cline{3-4}

& & \parbox{0.19\textwidth}{\small
The red-blue cartograms show the U.S. Presidential Election results in three different years.}

& 
\parbox{0.18\textwidth}
{
\hspace{-0.1cm}
\parbox{0.19\textwidth}{\small 
Which one of these was the closest election between the republicans (red) and the democrats (blue)?}
}
& & \\

\cline{3-4}

& & \parbox{0.19\textwidth}{\small
 Two separate US population cartograms of 1960 and 2010 are shown.}

& \parbox{0.18\textwidth}{\small
 What can you say about the trend in population growth?} & & \\

\hline
\end{tabular}

\caption{\small For each task, the last two columns show average completion time in seconds
 and error percentage for different cartogram types, along with the $F$ and $p$ values from
 ANOVA F-tests. The critical values of $F$ are $2.68$, $3.09$, and $3.99$ for analysis of 4, 3, and 2 algorithms, respectively. The bottom and top of the boxes and the blue band represent first quartile, third quartile and mean, respectively.
The upper and lower whiskers represnt the maximum and minimum values, respectively.
 The red line segments indicate statistically significant relationships, obtained using
 paired $t$-tests with Bonferroni correction. The critical values of $t$ are $2.81$, $2.52$ for pairwise comparison between 4 and 3 algorithms, respectively.}
  \label{fig:userstudy}
\label{tab:tasks}
\end{table*}

\section {Experiment}
\label{sec:experiment}


 We conduct a series of controlled experiments aimed
at producing a set of design guidelines for creating effective cartograms. We assess the effectiveness of our visualizations by performance (in terms of accuracy and completion time for visualization tasks) and subject reactions (attitude). 

\subsection{Hypotheses Formulation}
\label{sec:hypotheses}

Our hypotheses are informed by prior cartogram evaluations, perception studies,  and popular critiques of cartograms.
One of the most common criticisms is about shape distortion in cartograms, which makes it hard to recognize familiar geographic regions~\cite{Tobler04}. 
Dorling~\cite{dorling96} says ``A frequent criticism of cartograms is that even cartograms based upon the same variable for the same areas of a country can look very different.'' 
Tobler~\cite{Tobler04} reports ``It has been suggested that cartogram are difficult to use, although Griffin does not find this to be the case.'' 
Dent~\cite{dent1975}  suggests effective communication strategies such as providing an inset map and labeling.
With these comments in mind, in our experiment we added an undistorted map for the relevant tasks (\textit{locate}, \textit{detect change} \textit{find adjacency} and \textit{summarize}). 
We also labeled the regions for all tasks except \textit{locate} and \textit{recognize}, since labeling the regions for these two tasks would defeat the purpose of the tasks.
Before stating our hypotheses we note that we say that one cartogram type is ``better" than another for some task, when we expect quantitative differences (e.g., participants make fewer errors, or take less time) or qualitative differences (e.g., the participants prefer one over the other).


{\bf  H1}: For location tasks, contiguous and non-contiguous cartograms will be better than the other cartograms, as these two types preserve the relative position of regions~\cite{AKV15, Olson, cartogram-star}. Dorling cartograms will likely be better than rectangular cartograms.

{\bf H2:} For recognition tasks, non-contiguous cartograms are likely better than the rest since they preserve the original shapes~\cite{Olson}. (For recognizing the shape of a region we only test contiguous and
non-contiguous cartograms, because rectangular and Dorling cartograms replace the original shapes with rectangles and circles; testing shape recognizability would lead to predictably high errors and time).

{\bf H3:} For detecting change (whether a region has grown or shrunk in cartogram), and comparison of areas (size comparison, find top-$k$), contiguous cartograms are likely better than Dorling and rectangular cartograms, since the judgment of size of circles is difficult~\cite{teghtsoonian1965judgment}, and potentially large aspect ratios for rectangular cartograms can make the changes/comparisons difficult to perceive.
 
{\bf H4:} For finding adjacencies, contiguous and rectangular cartograms are likely better than the rest, because they preserve topology~\cite{AKV15, cartogram-star}, whereas non-contiguous and Dorling cartograms seem to be ill-suited for such tasks.

{\bf H5:} For summarizing the results and understanding data patterns, Dorling, non-contiguous and contiguous cartograms will work better than rectangular cartograms, as the first three types better preserve the map characteristics (location, shape and topology)~\cite{AKV15}.
With respect to subjective preferences, we expect that the participants in our study are likely to prefer contiguous
 and Dorling cartograms, as they are more frequently used.

\subsection{Participants}
 We recruited participants by sending email to students in selected classes at the University of Arizona: 
 
{\it ``We would like to invite you to take part in a research study to evaluate the usability of cartograms. A cartogram is a map in which some thematic mapping variable (e.g., population, income) is represented by the land area. The study takes 35--40 minutes: you will be asked to perform several tasks using cartograms and to compare different types of cartograms on a computer. All data will be collected anonymously and will be used for research purposes only. Modest compensation (\$10) will be provided for all participants. 
If you are interested, please find a convenient 1 hour time slot and provide your name and email address below.''}

The participants took part in the experiment one at a time, so that the experimenter could ensure that each participant understood the tasks at hand and had all their questions answered prior to starting the timed portion of the experiment. The participants were encouraged to ask as many questions as needed during the training session as well. 
All participants completed the experiment successfully, and no data was discarded.

Out of the 33 participants that took part in the study, 24 were male and 9 female; 23 between 18--25 years of age and 10 between 25--40; 9 listed high school, 12 listed undergrad, 8 listed Masters and
 4 listed PhD as their highest completed education level.
Familiarity with cartograms also differed:
14 participants were familiar with Dorling, 11 were familiar with contiguous, 8 with rectangular,  and
 3 with non-contiguous cartograms.

 Since some of our tasks require the subjects to identify
 regions highlighted with different colors, all participants were tested for color blindness using an \textit{Ishihara test}~\cite{Ishihara17}, and every participant passed the test. We used red and blue colors for highlighting.

\subsection {Test Environment}
We designed and implemented a simple application software that guided the participants through the experiment, provided task instructions and collected data about time and accuracy. 
The study was conducted using a computer (with i7 CPU 860 @ 2.80 GHz processor and 24 inch screen with 1600x900 pixel resolution), where the participants interacted with a standard mouse and keyboard to answer the questions. 
The experiment consisted of preliminary questions, a familiarity and initial ratings survey,  task-based questions, and preference and attitude questions.

\medskip
\noindent
\textbf{Preliminary questions:}
At the beginning, the participants filled out a standard human-subjects form confirming their participation in the experiment. They were also briefed about the purpose of the study: what cartograms are and what kind of tasks they will be asked to perform. The participants then completed some training tasks, familiarizing themselves with they software. 
Before proceeding to the next stage, each participant was given one more chance to ask questions and   were  told they would be able to take a break or leave the experiment whenever they wanted. All participants completed the experiment.

\medskip
\noindent
\textbf{Main experiment:} The main experiment had several stages:

\textit{Familiarity and initial rating:}
The participants reported their familiarity with each of the four cartogram types. For each type we showed   one example, along with a short description, and asked whether they were familiar with this particular type of cartogram.
We also asked the participants for an initial rating of the four cartogram types, using a Likert scale (excellent = 5, good = 4, average = 3, poor = 2, very poor =1).

\textit{Task-based questions:} 
For the task-based part of the study, the participants answered multiple choice questions about different visualizations, using all four types of cartograms under consideration, and showing different statistics for different countries (described in mode detail below). We recorded the number of correct and incorrect answers, as well as the time taken to provide the answers.

\textit{Preference and attitude study:} 
After all the tasks were completed, we asked the participants to choose one of the four cartogram types for another five questions. The goal of this set of questions was to help us detect whether the initial preferences might change after performing 67 timed tasks. For these five questions we were not interested in the time and error, but just in the choice that was made.

For the attitude study we adapted Dent's semantic differential technique~\cite{dent1975}. We used a rating scale between pairs of words or phrases that are polar opposites. There were five marks between these phrases and the participants selected the mark 
that best represented their attitudes for a given map and a given aspect. We used three aspects: general attitude about helpfulness and usability of the visualization, appearance, and readability.


\subsection{Datasets and Questions}

We evaluated four different types of cartograms, using seven types of tasks. In order to guard against potential bias introduced by only one or two datasets, we used three different maps (USA, Germany, Italy) and eleven different geo-statistical datasets. Specifically, for the first six tasks we used population and GDP of the USA, Germany and Italy from 2010. For \textit{summarize} tasks we used population of the USA in 1960 and 2010; GDP of Germany in 2010, crime rate in Italy, and three election results (2000, 2004, and 2008) in the USA. 

We used a within-subject experimental design. 
For each subject, questions were selected from all the cartogram types and all the tasks. To guard against adversary effects from the order of the questions, we took a random permutation of the questions for each subject. For each of the tasks, the participants worked with all three country maps (USA, Germany, Italy). 

In order to make a fair comparison we also wanted the participants to work with all four cartogram types for each task. Indeed, the participants worked with all four cartogram types for all the tasks, with two unavoidable exceptions. 
First, for \textit{recognize} tasks we used only contiguous and non-contiguous cartograms, since all the region shapes in Dorling and rectangular cartograms are circles and rectangles. Asking the participants to recognize the shape of a given region, when every region is a circle or a rectangle would be an unreasonably difficult challenge and might affect performance on other (and more meaningful) tasks.
Second, for \textit{detect change} we omit non-contiguous cartograms, since they use a different normalization of the areas than the other cartograms.
In particular, as described in Section~\ref{sec:types}, the size of a region in a non-contiguous cartogram is not directly related to the statistical data for that region, but it also depends on the distribution of the statistical data across all the regions. Thus, determining whether one region has grown or shrunk in a non-contiguous cartogram would be an unreasonably difficult challenge.

For each task, the questions were drawn from a pool of questions involving all possibly cartograms. 
Therefore, each participant answered 4 cartograms $\times$ 3 maps = 12 questions for four of the tasks (\textit{locate}, \textit{compare}), \textit{find top-$k$}, and \textit{find adjacency}). Since we evaluated only contiguous and non-contiguous cartograms for \textit{recognize}, this task involved 2 cartograms $\times$ 3 maps = 6 questions. Similarly for \textit{detect change} there were 3 cartograms $\times$ 3 maps = 9 questions. Finally for \textit{summarize}, where the participants compared and analyzed the overall data trends in the map, we used 4 different data sets: crime rate (arson) in Italy, GDP of Germany, population change (from1960 to 2010) in the USA, and Presidential election results in the USA. These four datasets were used on four different cartograms for each subject. In total, there were 4 tasks $\times$ 12 questions +  6 questions + 9 questions + 4 questions = 67 cartogram task-based questions. 
The order of the tasks, and the cartograms varied for each user.

\section{Results and Data Analysis}
\label{sec:results}

In this section, we report and analyze the results of our task-based quantitative experiment and qualitative experiment (subjective preferences and attitude study). Finally, we compare and contrast the metric-based data with the quantitative and quantitative data.


\subsection{Results of the Task-Based Study}

We use ANOVA $F$-tests with significance level $\alpha = 0.05$ to carry out the statistical analysis. The within-subject independent variables are the four cartogram types. The two dependent measures are the average completion times and error percentages by the participants, shown in the last two columns of Table~\ref{tab:tasks}. 
The null hypothesis is that the cartogram type does not affect completion times and error rates. 
 When the probability of the null hypothesis ($p$-value) is less than $0.05$ (or, equivalently the $F$-value is greater than the critical $F$-value, $F_{cr}$), the null hypothesis is rejected. For significance level
 $\alpha = 0.05$, the critical value of $F$ is  $F_{cr}=F_{0.05}(3,128)=2.68$ for all tasks except for recognize and detect change. For these two tasks the critical values are $F_{0.05}(1,64)=3.99$ and $F_{0.05}(2,96)=3.09$, respectively. 
 
 There is strong evidence for rejecting the null hypotheses in several cases; see Table~\ref{tab:tasks}. When the null hypothesis is rejected, paired $t$-tests are utilized for the post-hoc analysis, with Bonferroni correction on the significance level $\alpha = 0.05$. For each pair of cartogram types, we conclude that there is a significant difference in the mean completion time (respectively, mean error rate), if the computed $t$-value is greater than the critical $t$-value, $t_{cr}$.
 In pairwise comparison between 4 algorithms (i.e., 6 different pairs), the critical value of $t$ is $t_{cr}=t_{0.05/6}(32)=2.81$ (for all tasks except detect change and recognize). In pairwise comparison between 3 algorithms (i.e., 3 different pairs), the critical value of $t$ is $t_{cr}=t_{0.05/3}(32)=2.52$ (for detect change task). For the recognition task, only two algorithms are involved and hence a post-hoc analysis is not required.

\medskip
\noindent
\textbf{Hypothesis 1:}
H1 is based on the expectation that cartograms that preserve the relative position of the regions in the map facilitate \textit{locate} tasks. In particular, contiguous and non-contiguous cartograms should outperform the other two, with Dorling cartograms expected to be better than rectangular cartograms.
Indeed, there is strong evidence in support of this hypothesis, based on the results of the \textit{locate} task.
In particular, there are statistically significant differences (both completion times and error rates)
in performance between contiguous and rectangular cartograms, and between non-contiguous and rectangular cartograms. 

Dorling cartograms require significantly more time than non-contiguous cartograms, and are associated with significantly more errors than contiguous cartograms.
 There is also a statistically significant difference in the error rate for Dorling cartograms compared with rectangular cartograms, although the difference in completion times is not significant. In essence, the performance of the four types of cartograms varied as we expected, although in few cases, the differences were not statistically significant.

\medskip
\noindent
\textbf{Hypothesis 2:}
H2 is based on the expectation that non-contiguous cartograms should facilitate \textit{recognize} tasks, since they perfectly preserve the shapes of the regions from the geographic map.
Again, there is evidence in support of this hypothesis, based on the results of the \textit{recognize} task. In particular, 
there is a statistically significant difference in the error rates of contiguous and non-contiguous cartograms.
Moreover, the difference in errors is very large, at nearly a factor of four. 
Although there is no statistically significant difference for completion times, there are notable differences. For example, the range of time required for contiguous cartograms is much larger (5 - 45 seconds). Also note the bimodal distribution in the error plots for contiguous cartograms, with a peak at around 5\% error and another peak around 30\% error -- a different pattern from the unimodal distribution for non-contiguous cartograms, which peaks around 1\% error; see Table~\ref{tab:tasks}. 

One plausible explanation for the larger time range and the bimodal error distribution for contiguous cartograms, is that some participants took longer time than usual and sometimes found the correct answer, whereas others took little time and had little success finding the correct answer. While the average time is roughly the same time as for non-contiguous cartograms, the pattern is very different. Note that we intentionally did not evaluate Dorling and rectangular cartograms for  \textit{recognize} tasks, since recognizing the shape of a given region in a sea of circles or rectangles is impossible. Nevertheless, we can confidently say that non-contiguous cartograms are most suited for  \textit{recognize} tasks among the four types under consideration.

\medskip
\noindent
\textbf{Hypothesis 3:}
H3 is based on the expectation that contiguous cartograms should facilitate \textit{detect change} and \textit{compare} tasks, since these kinds of tasks are more difficult with circles and rectangles with possibly poor aspect ratios. There is partial evidence in support of this hypothesis, based on the three tasks used to test it: \textit{compare}, \textit{find top-$k$}, \textit{detect change}. 
Indeed for all three tasks the errors were the lowest in the contiguous cartogram setting. However, there were statistically significant results only in a subset of the possible pairs. 
In particular, there is a statistically significant difference in the error rates between contiguous and rectangular cartograms for all three tasks. 
 Even though the time spent was the lowest in the contiguous cartogram setting for two of the three tasks, there was statistical significance between contiguous and rectangular cartograms for only one task.
 
We used a relative difference in areas for the \textit{compare} task in the range (1.5, 4). We considered factors smaller than 1.5 too difficult and larger than 4 too easy. 
Although previous cognitive studies show that judgment of circle sizes is not very effective, in our study Dorling cartograms performed well for simple comparison between regions. This could be due to the fact that our \textit{compare} task was too easy (minimum ratio was 1.5), or because we did not ask the participants to estimate the size (area) of circles exactly, but rather to compare two circles and to find the circle with the larger area. 
For the more complex tasks of \textit{find top-$k$}, and \textit{detect change}, the error rates in Dorling cartograms are indeed significantly higher than contiguous cartograms, although there was no statistically significant difference in the time required.

\medskip
\noindent
\textbf{Hypothesis 4:}
H4 is based on the expectation that cartograms that preserve topology (i.e., contiguous and rectangular cartograms) would facilitate \textit{find adjacency} tasks. 
There is partial evidence in support of this hypothesis, based on the results for the \textit{find adjacency} task. Specifically, there is a statistically significant difference between the performance on contiguous and rectangular cartograms compared against Dorling and non-contiguous cartograms, in terms of error rates, although the same is not true for completion time. 

Note that for this task we provide an undistorted geographical map along with the cartogram, as suggested by Dent~\cite{dent1975} and Griffin~\cite{Gri83}. Despite this, the average error rates for non-contiguous (48.5\%) and Dorling cartograms (24.2\%) are much larger than the average error rates of rectangular (5\%) and contiguous cartograms (11.1\%). This implies that even in the presence of the original undistorted map, the cartograms which preserves topology significantly help the viewer finding the correct adjacency.

\medskip
\noindent
\textbf{Hypothesis 5:}
H5 is based on the expectation that Dorling, non-contiguous and contiguous cartograms should be better at showing geographic trends and patterns than rectangular cartogram, since they better preserve the map characteristics. There is partial evidence in support of this Hypothesis, based on the results of the \textit{summarize} task. In particular, the error rate for rectangular cartogram is the highest among all four cartograms. For both non-contiguous and Dorling cartograms, this difference in error rate is statistically significant. While the difference in error rate between contiguous and rectangular cartograms is not statistically significant, the error rate in rectangular cartograms is nearly twice that in contiguous cartograms. The completion time does not vary significantly among these cartograms, perhaps because this is a complex task where the participants spent significant time for each type. It is worth noting the wide distribution of errors and time for all four types. Participants took over 100 seconds to answer one \textit{summarize} question with rectangular and contiguous cartograms, while non-contiguous and Dorling required less than 75 seconds. 
All four cartograms yielded bimodal distributions of errors.

In general, the results of this part of the study show significant differences in performance (in terms of time and accuracy) between the four types of cartograms. As indicated by our hypotheses, different tasks seem better suited to different types of cartograms.
 Achieving perfection (with respect to minimum cartographic error, shape recognizability and topology preservation) in cartograms is difficult and no cartogram is equally effective in all three dimensions. Rectangular cartograms preserve adjacency relations, and that is reflected in the results. Non-contiguous cartograms maintain perfect shape, making the \textit{recognize} task easy, but the ``sparseness'' of the map makes it difficult to understand adjacencies. Dorling cartograms disrupt the adjacency relations but somewhat preserve the relative positions of regions, and are good at getting the ``big picture.'' Contiguous cartograms more or less preserve localities, region shapes, and adjacencies, and give the best performance for almost all the tasks. The familiarity with contiguous cartograms might play a role in this regard.

\subsection{Subjective  Preferences}

\begin{figure}[h]
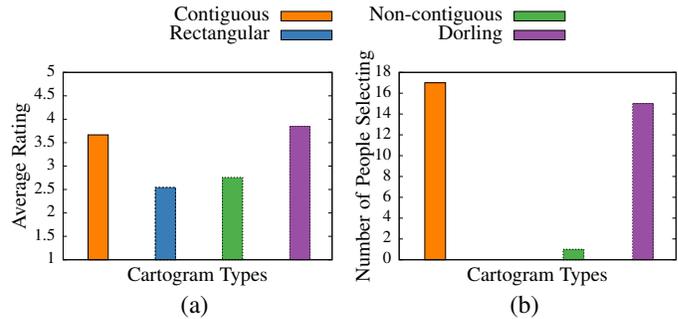

\centering
\scalebox{0.65}{\input{key_hor.tex}}\\
\scalebox{0.45}{\input{first-choice.tex}\input{last-choice.tex}}\\
\hspace{0.02\textwidth}(a)\hspace{0.22\textwidth}(b)
\caption{(a) Subjective cartogram ratings; (b) number of participants selecting a cartogram for  remaining tasks.}
\label{fig:first-choice}
\end{figure}

As described in Section~\ref{sec:experiment}, we asked the participants several preference questions in addition to the visualization tasks. At the beginning of the experiment, after introducing the different types of cartograms, the participants were asked to rate all four cartograms using a Likert scale (excellent = 5, good = 4, average = 3, poor = 2, very poor =1); see Fig.~\ref{fig:first-choice}(a). The results confirm our expectation that Dorling (average 3.84) and contiguous (3.66) are rated higher than non-contiguous (2.75) and rectangular (2.54).

\begin{figure*}[htb]
\centering
\scalebox{0.75}{\input{key_hor2.tex}}\\
\includegraphics[width=0.43\textwidth]{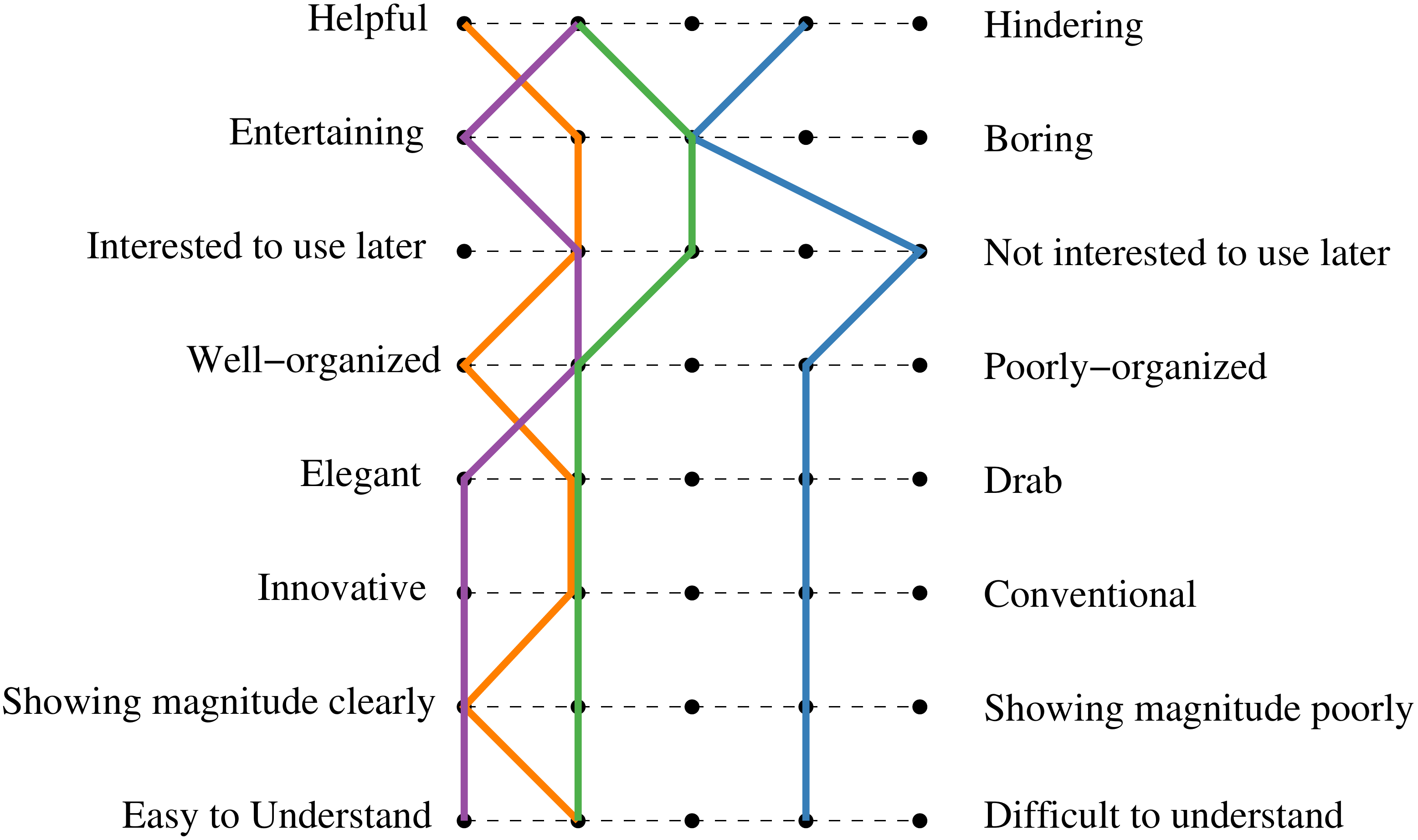}\hspace{.9cm}
\includegraphics[width=0.43\textwidth]{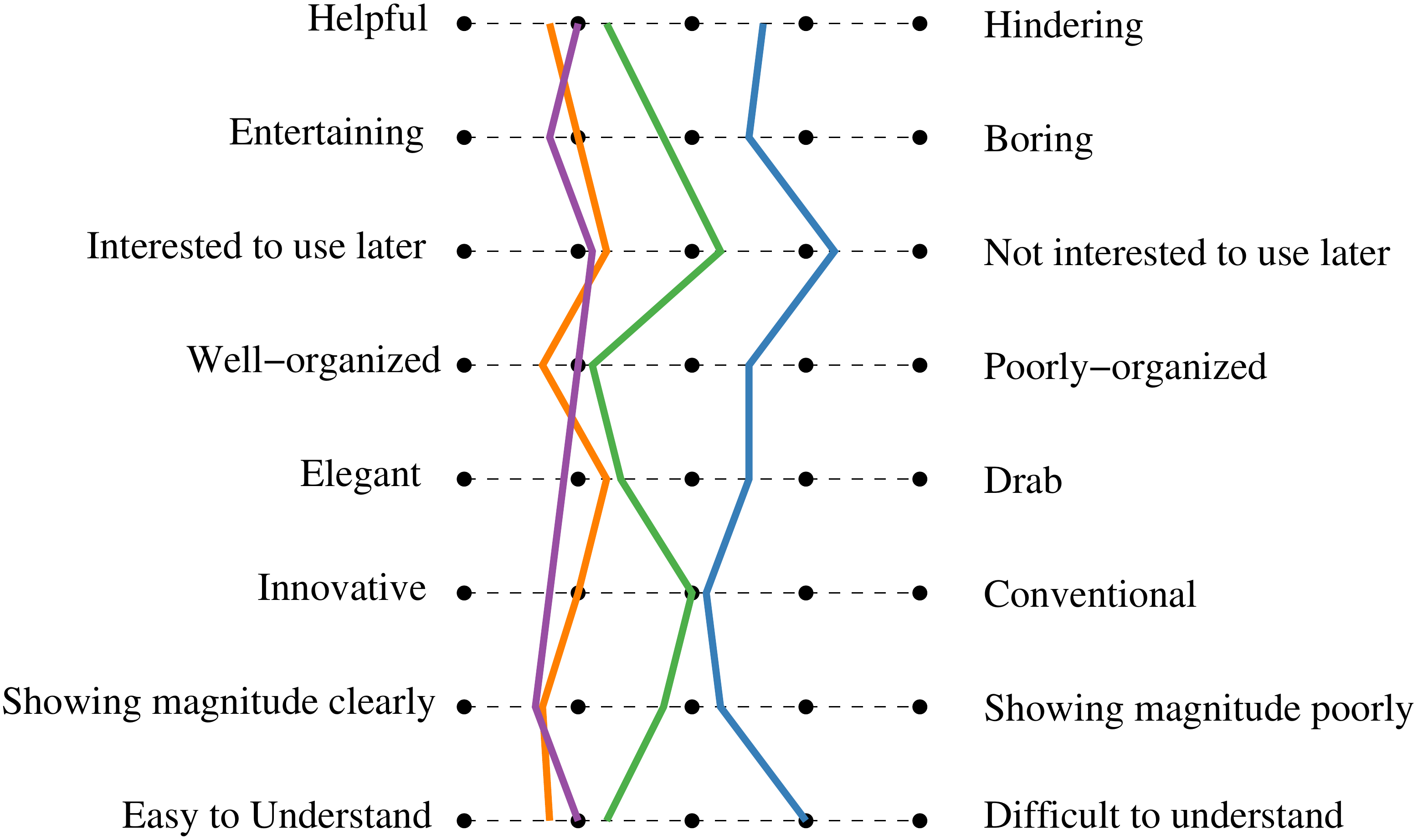}
\caption{
Attitude study of different cartograms by mode (left) and mean (right): contiguous and Dorling cartograms dominate.}
\label{fig:attitude}
\end{figure*}

After performing the visualization tasks, the participants were asked to select one of the four cartograms which would be used for an additional group of five questions. We asked this in order to test which cartograms were selected {\em after} performing many tasks and experiencing the different types of cartograms. The actual five questions were selected at random from the previous pool of questions, and the time and error rates for those five questions were not relevant. We were interested in the choices and in any changes from the preliminary ranking. Contiguous and Dorling cartograms remained the most preferred cartograms, although the order of the top two choices changes: out of 33 participants, 17 chose contiguous, 15 chose Dorling, 1 chose non-contiguous, and 0 chose rectangular; see Fig.~\ref{fig:first-choice}(b).  
In addition to the ease and efficiency in performing tasks with these two cartograms, the
 preference for contiguous and Dorling cartograms might partially be due to familiarity
 with these two cartograms in the news and on social media (10 participants reported that they are familiar with contiguous cartograms and 15 were familiar with
 Dorling cartograms, contrasted with 7 for rectangular cartograms and 2 for non-contiguous cartograms).

\subsection{Attitude Study}
As described in Section~\ref{sec:experiment}, we collected information about the attitude of the participants, which can be valuable as argued by Stasko~\cite{Stasko06}. In particular, at the end of the experiment, the participants were asked to rate the different cartogram types according to categories such as the helpfulness of the visualization, readability, and appearance, with a 
 rating scale between pairs of polar opposite words and phrases.
We considered the mode (most frequent response) and the mean (average response); see Fig.~\ref{fig:attitude}. This data also indicates a clear preference for contiguous and Dorling cartograms over the rest. The participants found contiguous cartograms to be helpful, well-organized and showing relative magnitude clearly, and Dorling cartograms to be entertaining, elegant, innovative, showing magnitude clearly, and easy to understand. The ``Interested to use later?'' choices also favor contiguous and Dorling cartograms.

\newcommand{\yes}{\textbf{\color{OliveGreen}H}}
\newcommand{\no}{\textbf{\color{red}L}}
\newcommand{\med}{\textbf{\color{YellowOrange}M}}

\subsection{Summary of All Results}

Table~\ref{tab:summary} summarizes the results of the metric-based and task-based analyses of all four cartogram types.
The results are aggregated in four dimensions. The first three dimensions aggregate results on the measures and tasks related to statistical accuracy, geographical accuracy and topological accuracy; while the last one illustrates each cartogram's effectiveness in showing the \textit{big picture}, i.e., trends, patterns, and outliers.
Considering the results in Table~\ref{tab:summary}, together with they subjective preferences and attitudes of participants, allows us to make several general observations.

\begin{table}[htb]
\centering
\begin{tabular}{|c||c|c|c|c||c|c|c|c|}

\hline

\multirow{2}{*}{\parbox{2cm}{\centering\vspace{0.5cm}Dimensions}} 
& \multicolumn{4}{c||}{{Metric-Based}} & \multicolumn{4}{c|}{{Task-Based}}\\

\cline{2-9}

& \rotatebox{90}{Cont} & \rotatebox{90}{Rect} & \rotatebox{90}{\hspace{-0.1cm}NCon} & \rotatebox{90}{Dor} & \rotatebox{90}{Cont} & \rotatebox{90}{Rect} & \rotatebox{90}{\hspace{-0.1cm}NCon} & \rotatebox{90}{Dor} \\

\hline\hline

{\small Statistical Accuracy} & \med & \no & \yes & \yes & \yes & \no & \yes & \yes \\

\hline

{\small Geographical Accuracy} 
& \med & \no & \yes & \med & \yes & \no & \yes & \med \\

\hline

{\small Topological Accuracy} & \yes & \yes & \no & \med & \yes & \yes & \no & \med \\

\hline

{\small Big Picture} &\multicolumn{4}{c||}{- -} & \med & \no & \yes & \yes \\

\hline
\end{tabular}

\caption{The result for metric-based and task-based analysis for all cartogram types. For metric-based analysis, {\yes}, {\med}, {\no} represent high, medium and low accuracy, respectively; 
 for task-based analysis, they represent high, medium and low performance, respectively}
\label{tab:summary}
\end{table}

Comparing the results of the metric-based and task-based analyses shows remarkable consistency in each of the dimensions: in each row of Table~\ref{tab:summary} a high ({\yes}) or medium ({\med}) accuracy in the metric-based evaluation corresponds to a high ({\yes}) or medium ({\med}) accuracy in the task-based evaluation. This indicates a consistency in how the different metrics and different tasks capture the three dimensions of cartogram design: topological accuracy, geographical accuracy and statistical accuracy.

Rectangular cartograms are a clear outlier and they should be used carefully. They performed sub-optimally in both the analysis of quantitative efficiency and in the qualitative subjective preference. This suggests that cartograms that severely distort region shapes and relative positions from the original map should be used very carefully. 
 A  promising compromise might be offered by rectilinear cartograms, such as that in Fig.~\ref{fig:red-blue}(b), where instead of a rectangle, a more complex rectilinear polygon represents each region, so that the region shapes and locations are preserved better. Mosaic cartograms are a recent practical method for generating such rectilinear cartograms~\cite{cano2015mosaic}.

Non-contiguous cartograms are good performers (many Hs) in both the metric-based and task-based evaluation, but they are not particularly appreciated by the participants (based on subjective preferences and attitude). Although these cartograms preserve perfect shape and relative positions for the regions, this lack of appreciation might be due to the loss of a feel of a map from the lack of contiguity. Further, some regions become too small to recognize and overall there is more white space. 
One possible way to mitigate this is to compromise the perfect relative position by moving the regions to allow for them to scale up without overlapping and reduce the unused space.

Contiguous cartograms and Dorling are good performers (Ms and Hs) in both the metric-based and task-based evaluations; they are also well liked (subjective preferences and attitude).


\newcommand{\plotStudyB}[1]
{
	\parbox{\textwidth}
	{
		\centering
		\includegraphics[width=0.24\textwidth]{#1-familiar.pdf}
		\includegraphics[width=0.24\textwidth]{#1-gender.pdf}
		\includegraphics[width=0.24\textwidth]{#1-education.pdf}
		\includegraphics[width=0.24\textwidth]{#1-Age.pdf}
	}
}

\newcommand{\plotStudyG}[1]
{
	\parbox{0.35\textwidth}
	{
		\hspace{-0.4cm}
		\includegraphics[width=0.35\textwidth]{plots-split-bean/#1-gender.pdf}
	}
}

\newcommand{\plotStudyF}[1]
{
	\parbox{0.35\textwidth}
	{
		\hspace{-0.4cm}
		\includegraphics[width=0.35\textwidth]{plots-split-bean/#1-familiar.pdf}
	}
}

\begin{figure*}[htb]
\centering
\parbox{\textwidth}
{\plotStudyB{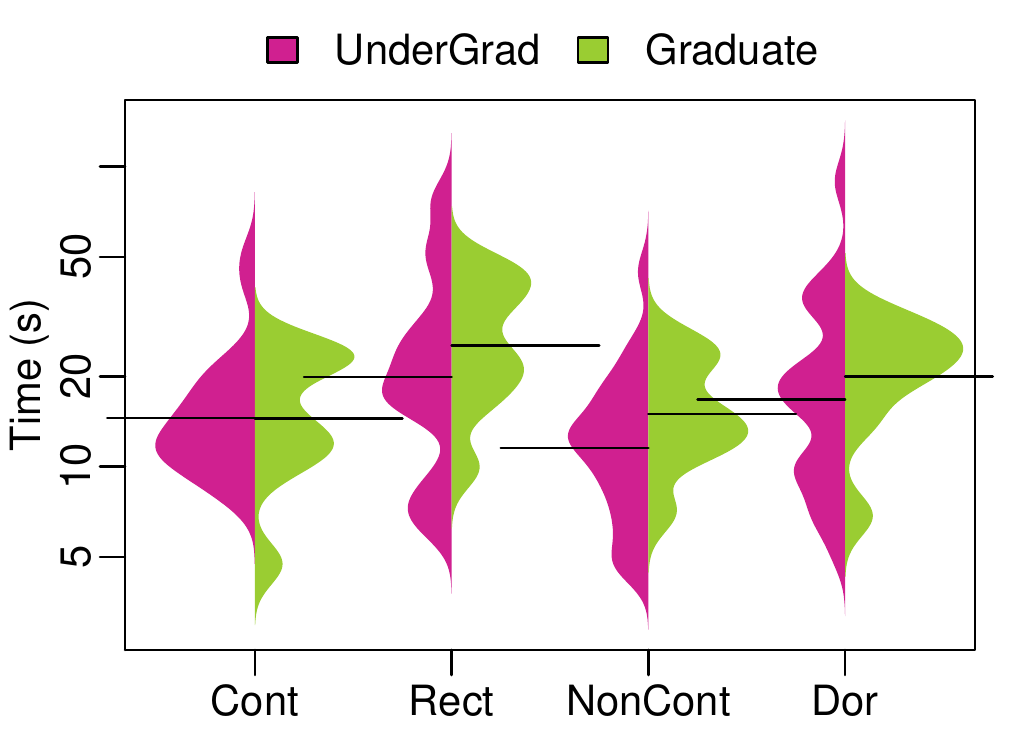}}\\(a) Locate\\

\parbox{\textwidth}
{\plotStudyB{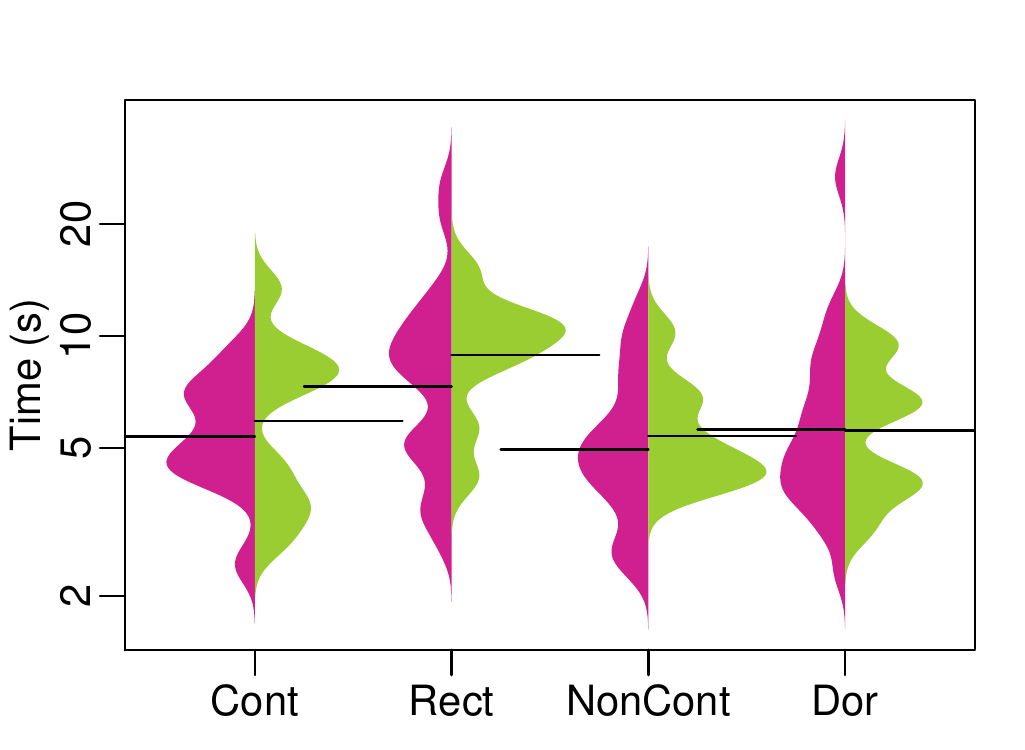}}\\(b) Compare\\

\parbox{\textwidth}
{\plotStudyB{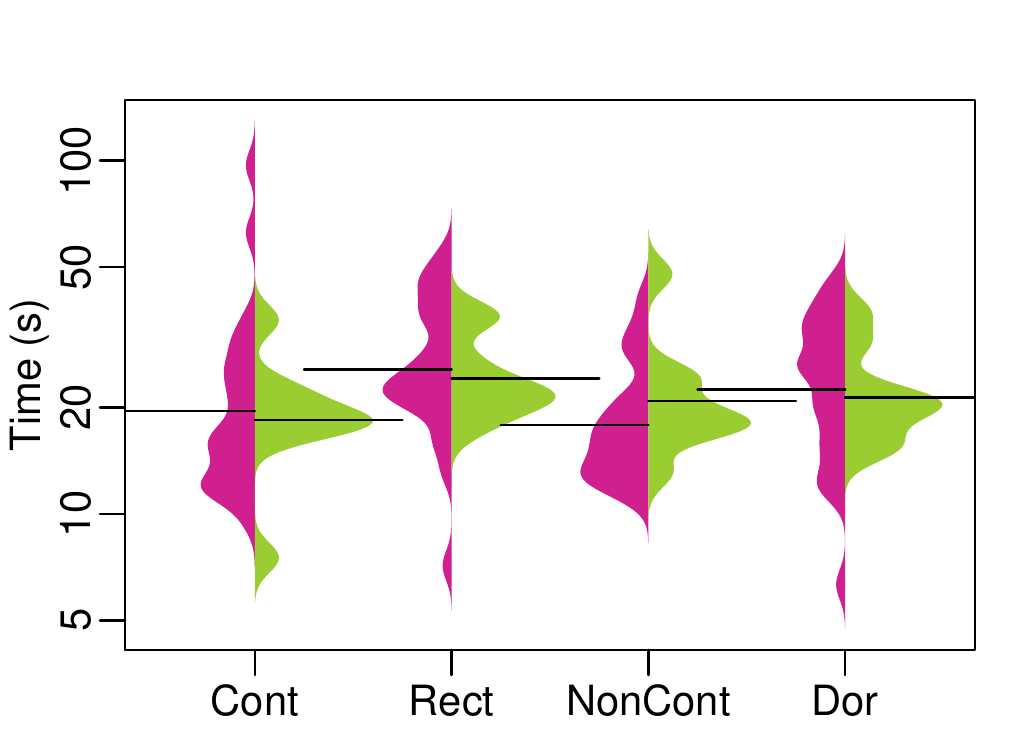}}\\(c) Find Adjacency\\

\parbox{\textwidth}
{\plotStudyB{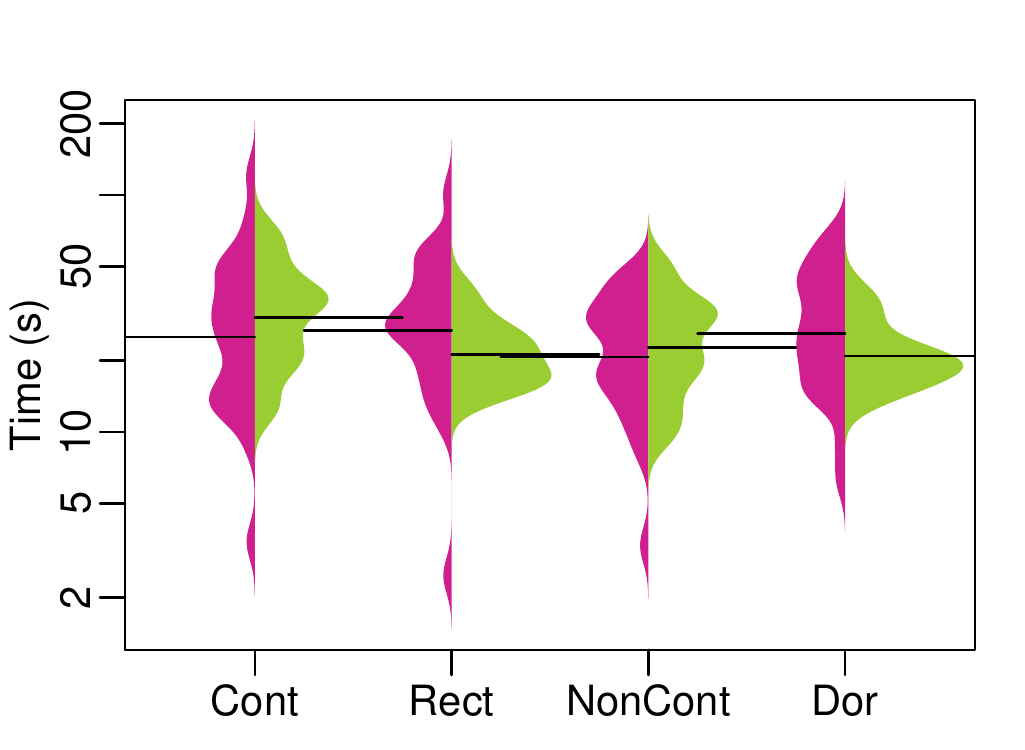}}\\(d) Summarize

\caption{Average completion times in seconds for different tasks on different cartograms by different demography groups. The first columns show completion times by by participants familiar or not familiar with the corresponding cartogram type; the second columns show comepletion times by female and male participants; the third columns by participants with undergraduate and graduate education level; and the fourth columns by participants under and over the age of 25years. Different rows show plots for different tasks: (a) locate, (b) compare, (c) find adjacency, and (d) summarize. The solid line on each side of each plot represents the mean completion time for the respective group of participants.}
\label{fig:demographics-time}
\end{figure*}

\begin{figure*}[htb]
\centering
\parbox{\textwidth}
{\plotStudyB{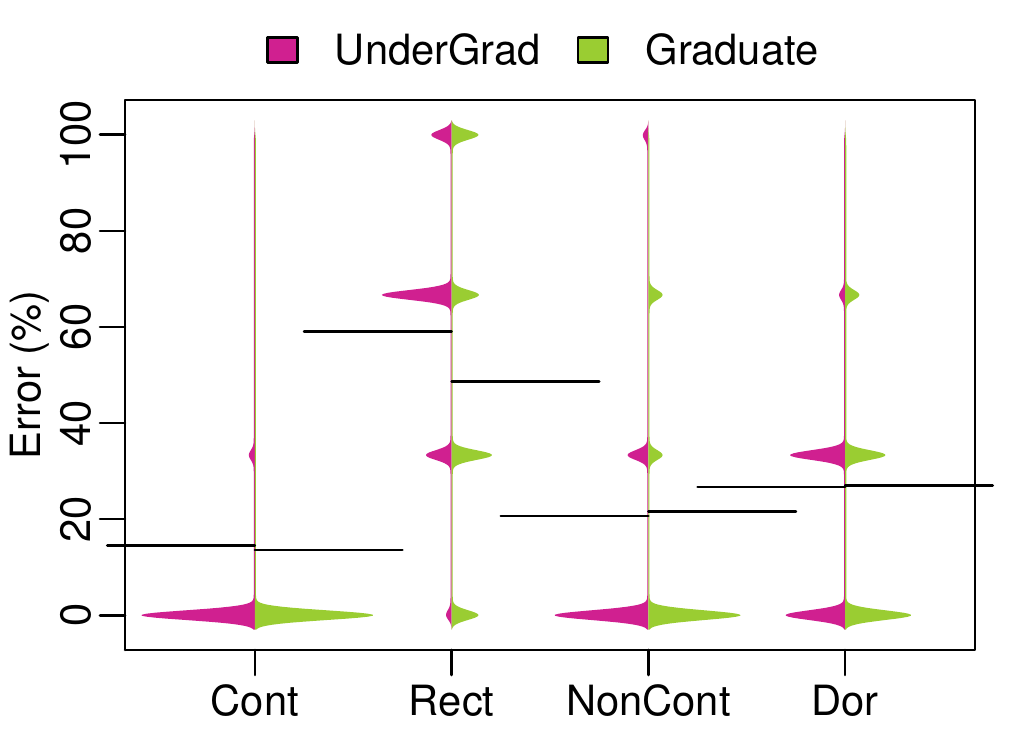}}\\(a) Locate\\

\parbox{\textwidth}
{\plotStudyB{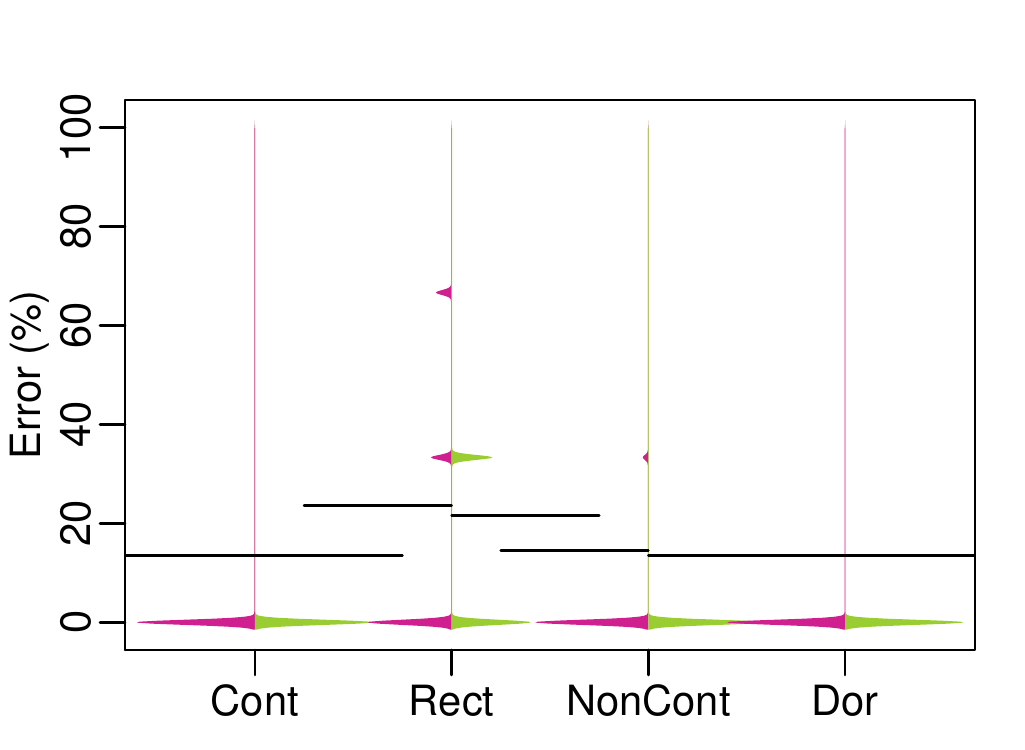}}\\(b) Compare\\

\parbox{\textwidth}
{\plotStudyB{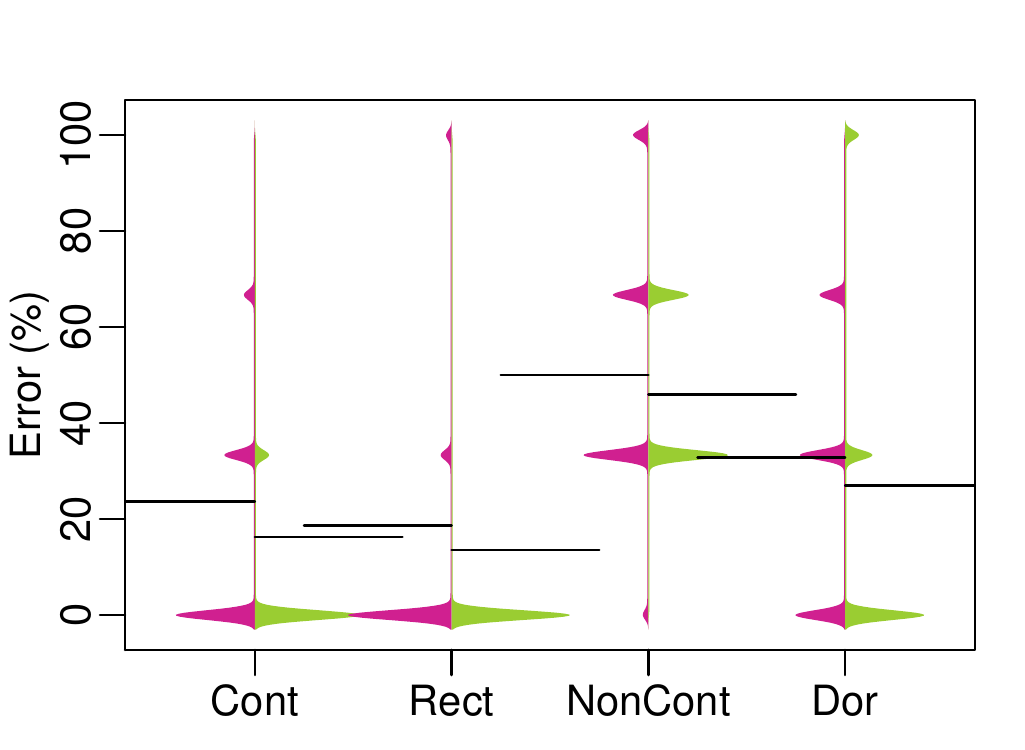}
}\\(c) Find Adjacency\\

\parbox{\textwidth}
{\plotStudyB{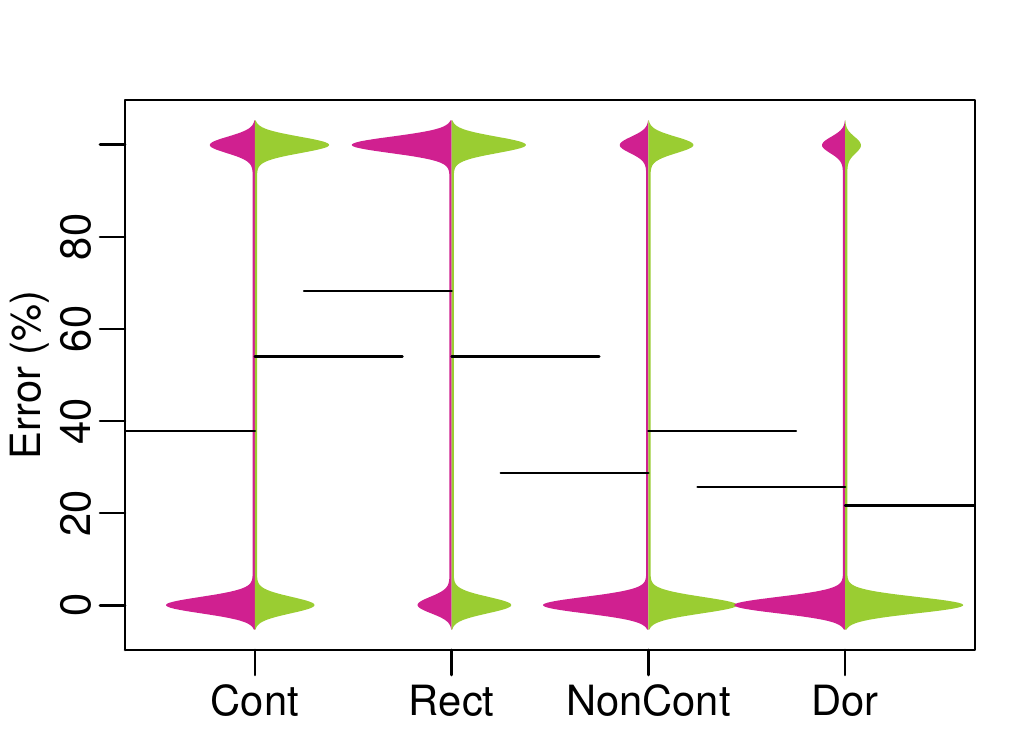}}\\(d) Summarize

\caption{Average error percentages for different tasks on different cartograms by different demography groups. The first columns show error percentages by by participants familiar or not familiar with the corresponding cartogram type; the second columns show error percentages by female and male participants; the third columns by participants with undergraduate and graduate education level; and the fourth columns by participants under and over the age of 25 years. Different rows show plots for different tasks: (a) locate, (b) compare, (c) find adjacency, and (d) summarize. The solid line on each side of each plot represents the mean error percentage for the respective group of participants.}
\label{fig:demographics-error}
\end{figure*}

\section{Demographic Analysis}

In this section, we consider how participants of different age groups, gender, and education levels perform in the study. In particular, our goal here is to find out how people with different background make sense of geographic data using different cartogram types. For this demographic analysis we selected a subset of the seven tasks from the study. These tasks measure all three dimensions of cartogram design: topological accuracy (\textit{find adjacency}), geographical accuracy (\textit{locate}), statistical accuracy (\textit{compare}), as well as composite tasks (\textit{summarize}).

We analyzed our task-based results as well as the subjective ratings in the context of different demographic groups: participants who are familiar and who are not familiar  with a particular cartogram type, male and female participants, participants under and over the age of 25, and undergraduate and graduate students. We discuss several interesting findings; see Figs.~\ref{fig:demographics-time},~\ref{fig:demographics-error},~and~\ref{fig:rating-demo}.

\subsection{Task-Based Performance of Demographic Groups}

 \textbf{Familiarity affects performance.} At the beginning of the study, we collected data about the familiarity of the participants with the four cartogram types. We analyzed the impact of familiarity with cartograms on the completion time and error rate; see Figs.~\ref{fig:demographics-time}~and~\ref{fig:demographics-error}. Subjects who were familiar with cartograms took significantly longer time to perform the tasks (the significance was tested with Welch's $t$-test), while the error rates seem not to be affected.
This seems counter-intuitive, as we expected participants familiar with a particular cartogram type should make fewer errors and be faster in their response.
One possible explanation is that familiarity is associated with deeper engagement: participants familiar with a cartogram type might have been more interested and engaged in the visualization.

\noindent
\textbf{Female participants were more accurate.}
We did not expect gender of the participants to be a factor in the accuracy of performing the tasks. However, in our study, female participants seem to be more accurate in most tasks involving contiguous, Dorling and non-contiguous cartograms (for contiguous cartograms, the difference in accuracy between the two groups is statistically significant (using Welch's $t$-test), but there is no such pattern for rectangular cartograms.  Completion times are not significantly different for male and female participants. 

\noindent
\textbf{Age and education did not affect performance.}
We considered the possibility that older participants and participants with higher education level might perform better, since they are likely to be more familiar with more cartograms and maps~\cite{monmonier2014lie}.
However, we did not find significant differences for different age groups and education levels. One possible explanation is that by using three different maps (USA, Germany, Italy) and a within-subject experiment design, the participants were not aided by knowledge of a particular map or cartogram type.

\subsection{Subjective Preferences of Demographic Groups}

\begin{figure*}[htb]
\centering
\parbox{\textwidth}
{
\centering
\includegraphics[width=0.24\textwidth]{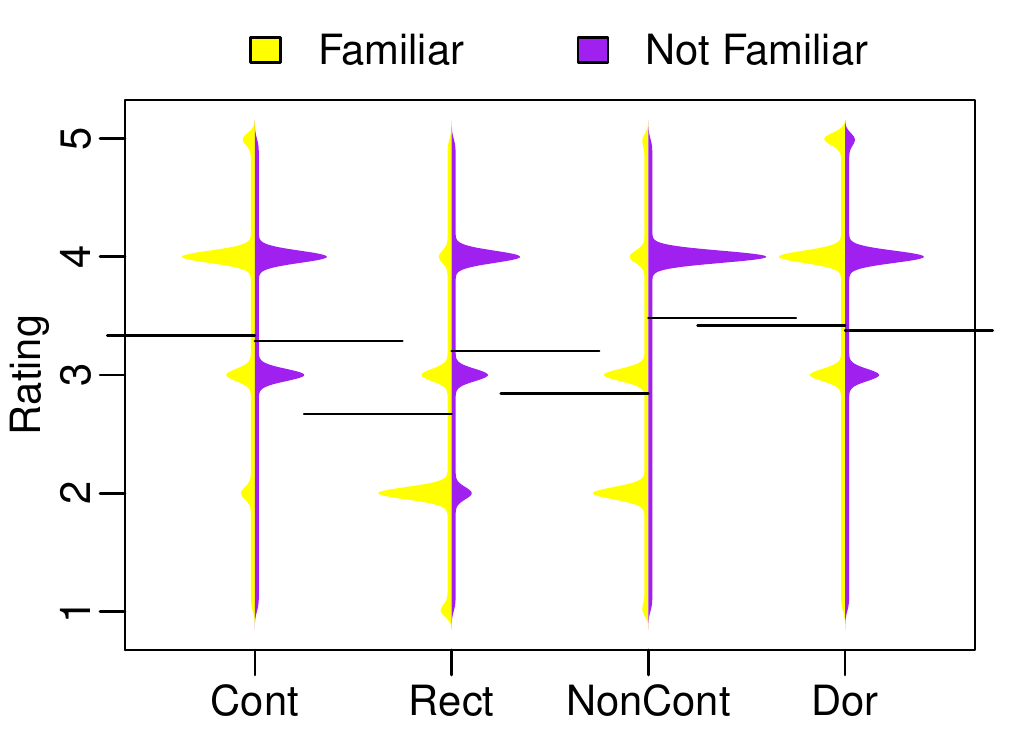}
\includegraphics[width=0.24\textwidth]{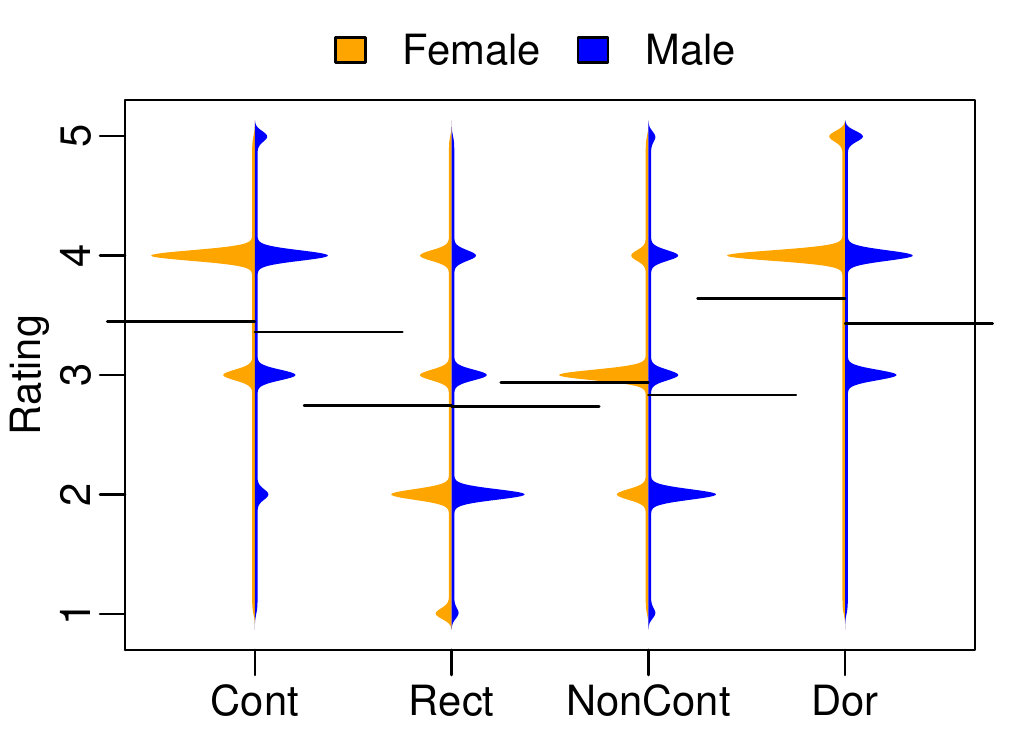}
\includegraphics[width=0.24\textwidth]{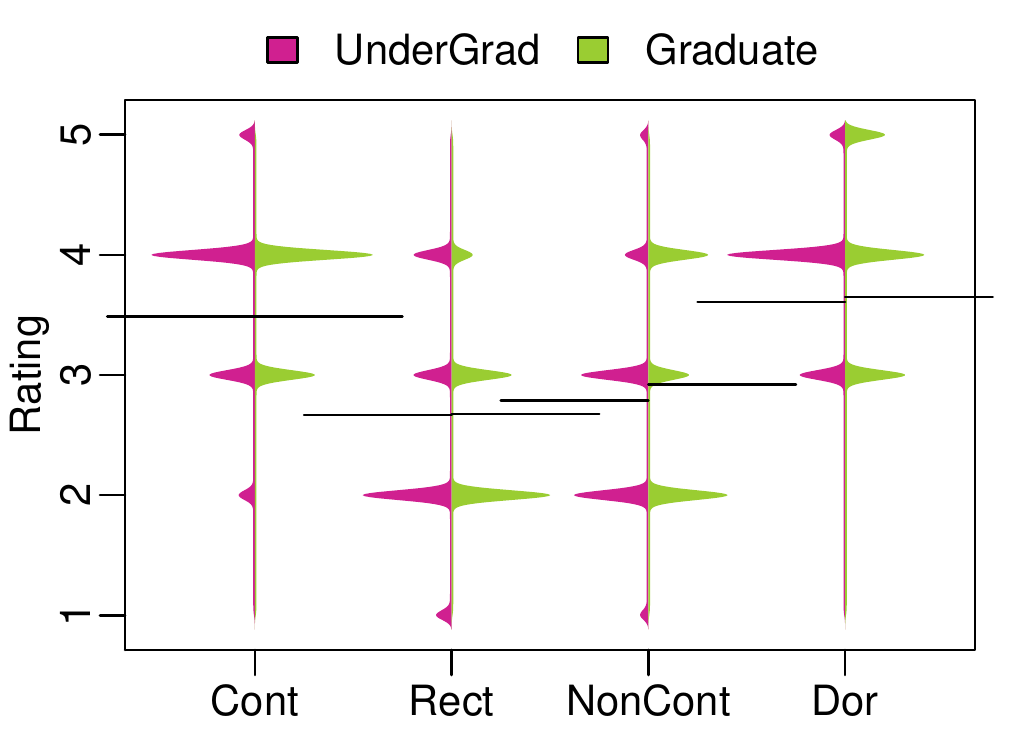}
\includegraphics[width=0.24\textwidth]{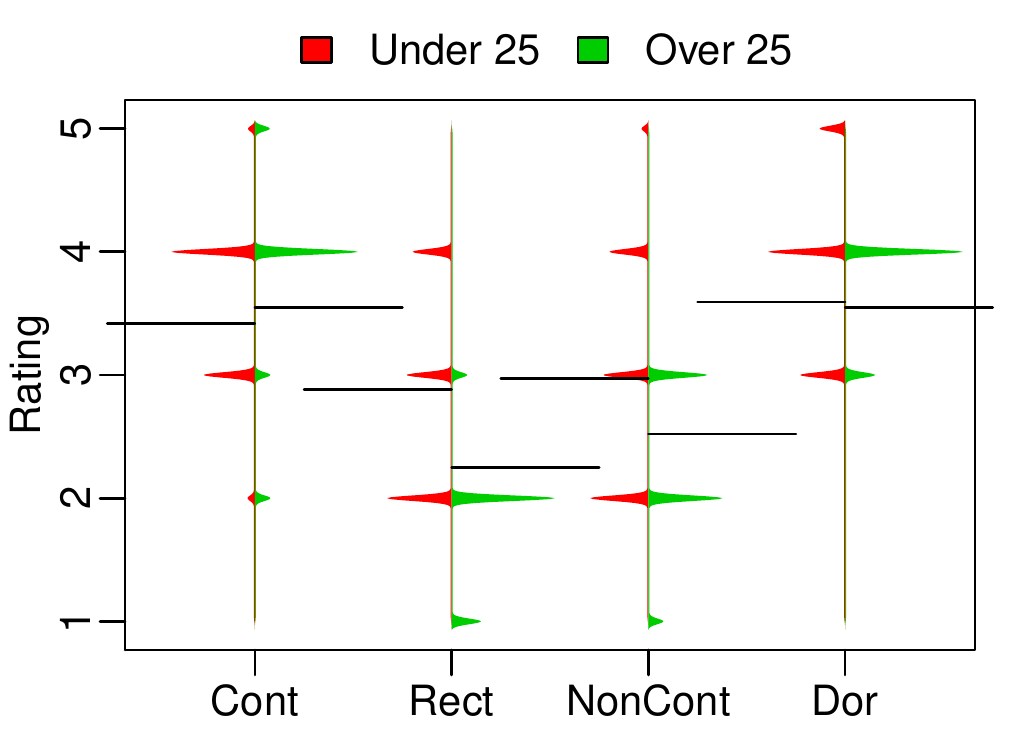}
}
\caption{Subjective ratings of different cartograms by participants familiar and not familiar with the cartogram type, by female and male participants, by participant with undergraduate and graduate education level, and by participants under and over the age of 25 years.}
\label{fig:rating-demo}
\end{figure*}

 \textbf{Female participants gave higher ratings.} Female participants rated all cartogram types, except rectangular cartograms, higher than their male counterparts. In particular, there is a strong indication (using  Welch's $t$-test) that female participants prefer Dorling cartograms more than the male participants; see Fig.~\ref{fig:rating-demo}(left).  Once possible explanation could come from earlier findings that round, circular shapes are preferred over sharp, angular shapes~\cite{silvia2009people, bertamini2016observers}.

\noindent
\textbf{Familiarity affected preferences.}
We anticipated that familiarity with a particular cartogram type might make this type more liked.
The participants gave similar ratings to unfamiliar cartogram types (around 3.5 on average), but different ratings to cartogram types they were familiar with. In the subjective ratings, contiguous and Dorling cartograms clearly outperform non-contiguous and rectangular cartograms. However, a closer look shows something interesting about rectangular and non-contiguous cartograms. Participants who were familiar with these two types of cartograms rated them lower than those who were unfamiliar; see Fig.~\ref{fig:rating-demo}. This is consistent with the choices made at the end of the study. After performing 67 tasks, all participants were familiar with all cartogram types, but hardly any participant chose rectangular or non-contiguous cartograms for the final 5 tasks.

\section{Discussion and Design Implication}

Cartograms are good at summarizing data and showing broader trends and patterns, as shown in early research~\cite{Krauss_ms, Hui}. While partially confirming some of these results, our study also identified significant differences in performance between different cartogram types and different tasks. This is relevant as new cartogram types continue to be created~\cite{cano2015mosaic} and identifying difficult tasks for specific cartogram types can lead to improvements in design.

\subsection{So, Which Cartogram is Best?}

\begin{figure}[h]
\centering
\includegraphics[width=0.3\textwidth, angle=270]{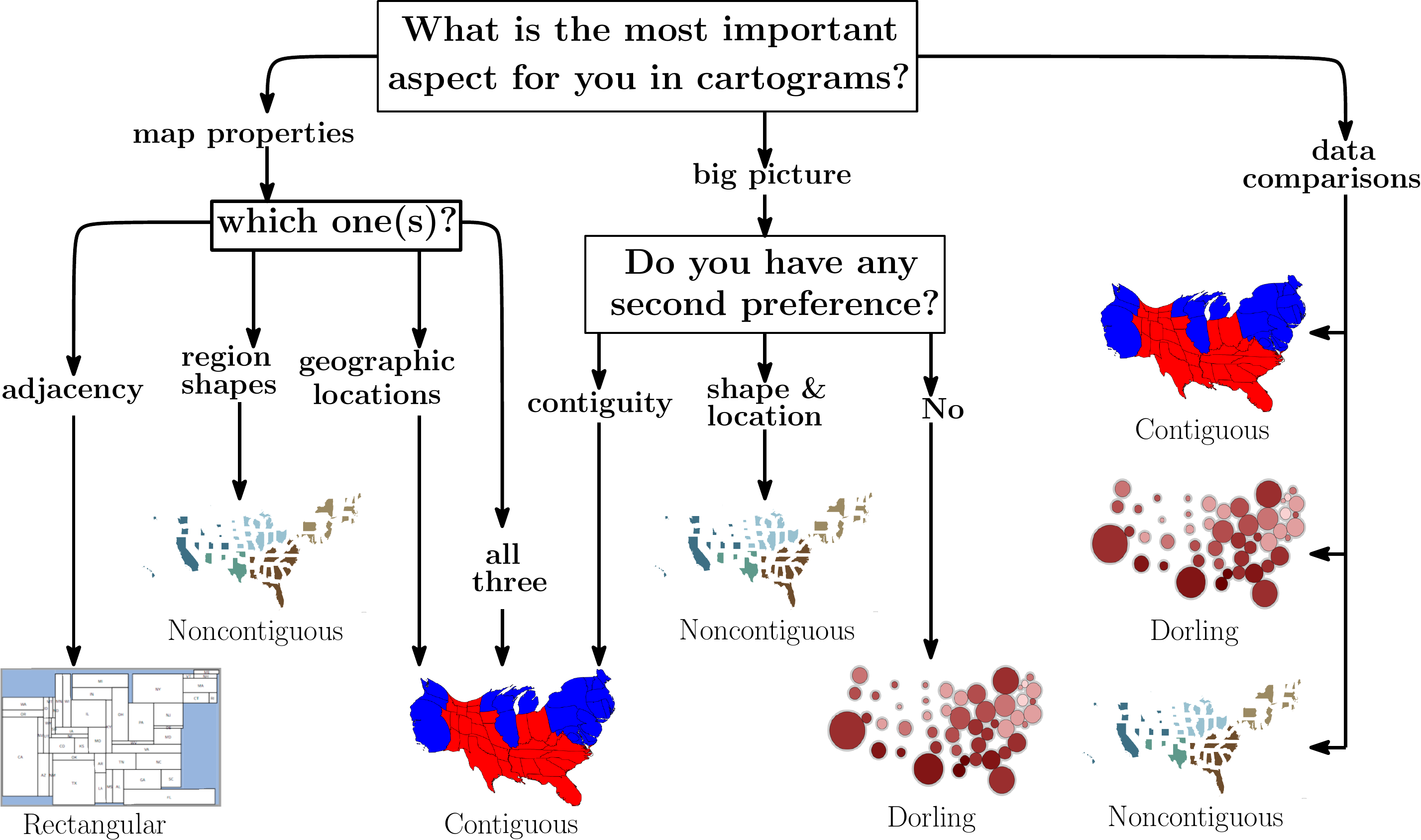}
\caption{Cartogram type flow diagram.}
\label{fig:picker}
\end{figure}

The choice of cartogram type should take into account the expected tasks.
All cartogram types, except rectangular, performed well in tasks involving analyzing and comparing trends, with Dorling cartograms giving the best results. The reason might be that the simple circular shapes convey the data pattern easily, whereas the distortion in shape and size for other cartograms distract the viewers. 
When the geographic locations and adjacencies are important aspects, and the required map-reading  is more detailed, contiguous cartograms might be more suitable. This seems to be the case for tasks, such as \textit{locate, find top-$k$,} and \textit{detect change}.
On the other hand,  rectangular cartograms work well if adjacency relations are important, and having a simple schematic representation is useful. For comparison of polygons, contiguous, non-contiguous, and Dorling all work equally well. 
We summarize these observations in a flowchart that could be used to guide the choice of a cartogram for a particular application; see Fig.~\ref{fig:picker}.

The choice of cartogram type should also take into account the type of map being shown. Countries with few regions, such as Italy and Germany, are easier to schematize, while still preserving the general outlines. Similarly, most of their regions are on the periphery, making it easier to shrink or grow individual regions. Countries with more regions (and more landlocked regions) are more difficult to deal with.

\subsection{Design Improvements by Interaction}
One of the design implications from this study is that simple interaction techniques might mitigate some cartogram shortcomings. 
For example, to reduce the effect of misinterpretation associated with area perception, exact values can be shown using mouse-over/tool-tip labels. Some interactive web visualizations already provide such features~\cite{NYT_O}. 
Non-contiguous cartograms  perfectly preserve region shapes and geographic locations,
 and their performance is good for almost all the tasks, with the clear exception of
 finding adjacencies. 
 Note, however, that the high errors for non-contiguous and Dorling cartograms on tasks such as  \textit{find adjacency} can be remedied by another simple interaction, such as mouse-over highlighting of all neighbors.
For cartograms where identifying a particular region by its shape is difficult (e.g., Dorling, rectangular), link-and-brush type highlighting of the corresponding region in a linked geographic map might alleviate the problem.  Such interactions, together with interactions that show exact values with mouse-over/tool-tip labels, will likely lead to improved performance for most cartogram types.

\subsection{Limitations}
We limited ourselves to one representative from each of four major types of cartograms. There are other types of cartograms and even more variants thereof (e.g., over a dozen contiguous cartograms) that we did not consider. Similarly, while attempting to cover the spectrum of possible cartogram tasks, we limited ourselves to a particular subset of tasks and particular choices for the task settings. There are numerous limitations when considering the types of possible geographic maps (e.g., more countries, continents, or even synthetic maps) and the relationship between the original geographic area and the statistical data shown (e.g., extreme area changes, moderate area changes, insignificant area changes).
Since each participants met with the experimenter in person, we had a small number of participants and not as wide a spread over age and background knowledge. Despite such limitations, we believe the results of our study will be of use.

\section {Conclusion}

We described a thorough evaluation of four major types of cartograms, going beyond time and error by synthesizing metrics-based, task-based, and subjective evaluations. The results show significant differences between cartogram types and provide insights about the effectiveness of the different types for different tasks. Given the popularity of cartograms in representing geo-spatial data, we believe that cartograms should be studied more carefully. While it is unlikely that a single evaluation study will be complete and will cover all possible issues, we feel that our work can be a useful starting point, while providing directions for future cartogram studies. We provide all details about this study (e.g., datasets, exact questions, answers, statistical analysis) available online at \url{http://cartogram.cs.arizona.edu/evaluations.html}.

A great deal of interesting future work remains. 
Cartograms are convenient tools for learning; and they are used in textbooks, for example, to teach middle-school and high-school students about global demographics and human development~\cite{Class2}. It would be worthwhile to study the effect of different cartogram types on engagement in the context of learning. Enjoyment is a concept related to engagement and while enjoyment is extensively studied in psychology and recently of interest in visualization there is little work in the context of cartograms.
Intuitively, it seems clear that being engaged with a visualization, enjoying it, and having fun can be beneficial, especially in the context of learning. 
Similarly, memorability (both in the context of recognition, e.g., ``have you seen this visualization before?'' and recall of data, e.g., ``can you retrieve data from memory about a visualization you have seen before?'') is relevant for cartograms and not well studied yet.



\begin{thebibliography}{10}

\bibitem{WorldMapper}
Worldmapper.
\newblock \url{http://www.worldmapper.org/}.

\bibitem{NYT_04}
{Election} 2004 -- {NYTimes}.
\newblock
  \url{http://www.nytimes.com/packages/html/politics/2004_ELECTIONRESULTS_GRAPHIC/},
  2004.

\bibitem{NYT_O}
{A Map of Olympic Medals -- NYTimes}.
\newblock
  \url{http://www.nytimes.com/interactive/2008/08/04/sports/olympics/20080804_MEDALCOUNT_MAP.html},
  2008.

\bibitem{zackary_blog}
indiemaps.com/blog $>>$ cartogram design.
\newblock \url{http://goo.gl/Loq7q7}, 2008.

\bibitem{NRC}
Europe goes to the polls.
\newblock \url{http://vorige.nrc.nl/international/article2238466.ece}, 2009.

\bibitem{Guar}
How to visualise social structure, {The Guardian}.
\newblock
  \url{http://www.theguardian.com/news/datablog/2012/jul/24/danny-dorling-visualise-social-structure},
  2012.

\bibitem{LAT12}
{U.S.} election results -- 2012 election -- {Los Angeles Times}.
\newblock
  \url{http://graphics.latimes.com/2012-election-results-national-map/}, 2012.

\bibitem{NYT16}
{A Map of Rio Olympic Medals -- NYTimes}.
\newblock
  \url{http://www.nytimes.com/interactive/2016/sports/olympics/rio-olympics-2016-medals-results.html},
  2016.

\bibitem{AKV15}
M.~J. Alam, S.~G. Kobourov, and S.~Veeramoni.
\newblock Quantitative measures for cartogram generation techniques.
\newblock {\em Computer Graphics Forum}, 34(3):351--360, 2015.

\bibitem{bertamini2016observers}
M.~Bertamini, L.~Palumbo, T.~N. Gheorghes, and M.~Galatsidas.
\newblock Do observers like curvature or do they dislike angularity?
\newblock {\em British Journal of Psychology}, 107(1):154--178, 2016.

\bibitem{BERTIN83}
J.~Bertin.
\newblock {\em Semiology of Graphics: Diagrams, Networks, Maps}.
\newblock The University of Wisconsin Press, Madison, 1983.

\bibitem{BM13}
M.~Brehmer and T.~Munzner.
\newblock A multi-level typology of abstract visualization tasks.
\newblock {\em {IEEE} Transaction on Visualization and Computer Graphics},
  19(12):2376--2385, 2013.

\bibitem{BSV12}
K.~Buchin, B.~Speckmann, and S.~Verdonschot.
\newblock Evolution strategies for optimizing rectangular cartograms.
\newblock In {\em Geographic Information Science}, volume 7478 of {\em Lecture
  Notes in Computer Science}, pages 29--42, 2012.

\bibitem{cano2015mosaic}
R.~G. Cano, K.~Buchin, T.~Castermans, A.~Pieterse, W.~Sonke, and B.~Speckmann.
\newblock Mosaic drawings and cartograms.
\newblock {\em Computer Graphics Forum}, 34(3):361--370, 2015.

\bibitem{cleveland1982judgments}
W.~S. Cleveland, C.~S. Harris, and R.~McGill.
\newblock Judgments of circle sizes on statistical maps.
\newblock {\em Journal of the American Statistical Association},
  77(379):541--547, 1982.

\bibitem{cleveland1984graphical}
W.~S. Cleveland and R.~McGill.
\newblock Graphical perception: Theory, experimentation, and application to the
  development of graphical methods.
\newblock {\em Journal of the American statistical association},
  79(387):531--554, 1984.

\bibitem{BMS10}
M.~de~Berg, E.~Mumford, and B.~Speckmann.
\newblock {Optimal BSPs and Rectilinear Cartograms}.
\newblock {\em International Journal of Computational Geometry \&
  Applications}, 20(2):203--222, 2010.

\bibitem{dent1975}
B.~D. Dent.
\newblock Communication aspects of value-by-area cartograms.
\newblock {\em The American Cartographer}, 2(2):154--168, 1975.

\bibitem{dorling96}
D.~Dorling.
\newblock {\em {Area Cartograms: Their Use and Creation}}, volume~59 of {\em
  Concepts and Techniques in Modern Geography}.
\newblock University of East Anglia, 1996.

\bibitem{DCN85}
J.~A. Dougenik, N.~R. Chrisman, and D.~R. Niemeyer.
\newblock An algorithm to construct continuous area cartograms.
\newblock {\em The Professional Geographer}, 37(1):75--81, 1985.

\bibitem{GN04}
M.~Gastner and M.~Newman.
\newblock Diffusion-based method for producing density-equalizing maps.
\newblock In {\em The National Academy of Sciences of the USA}, volume 101,
  pages 7499--7504, 2004.

\bibitem{Gri83}
T.~Griffin.
\newblock Recognition of area units on topological cartograms.
\newblock {\em The American Cartographer}, 10(1):17--29, 1983.

\bibitem{JM10}
J.~Heer and M.~Bostock.
\newblock Crowdsourcing graphical perception: using mechanical turk to assess
  visualization design.
\newblock In {\em {ACM SIGCHI} Conference on Human Factors in Computing
  Systems}, pages 203--212. ACM, 2010.

\bibitem{hkps04}
R.~Heilmann, D.~Keim, C.~Panse, and M.~Sips.
\newblock Recmap: Rectangular map approximations.
\newblock In {\em IEEE Symposium on Information Visualization (InfoVis'04)},
  pages 33--40, 2004.

\bibitem{HK98}
D.~H. House and C.~J. Kocmoud.
\newblock Continuous cartogram construction.
\newblock In {\em IEEE Visualization}, pages 197--204, 1998.

\bibitem{Ishihara17}
S.~Ishihara.
\newblock Tests for color-blindness.
\newblock 1917.

\bibitem{kaspar2013empirical}
S.~Kaspar, S.~Fabrikant, and P.~Freckmann.
\newblock Empirical study of cartograms.
\newblock In {\em International Cartographic Conference}, volume~3, 2011.

\bibitem{KNPS03}
D.~Keim, S.~North, C.~Panse, and J.~Schneidewind.
\newblock Visualizing geographic information: {VisualPoints vs. CartoDraw}.
\newblock {\em Information Visualization}, 2(1):58--67, 2003.

\bibitem{KNP04}
D.~A. Keim, S.~C. North, and C.~Panse.
\newblock {CartoDraw: A Fast Algorithm for Generating Contiguous Cartograms}.
\newblock {\em IEEE Transactions on Visualization and Computer Graphics},
  10(1):95--110, 2004.

\bibitem{KPN05}
D.~A. Keim, C.~Panse, and S.~C. North.
\newblock Medial-axis-based cartograms.
\newblock {\em IEEE Computer Graphics and Applications}, 25(3):60--68, 2005.

\bibitem{Krauss_ms}
M.~R.~D. Krauss.
\newblock {\em The relative effectiveness of the noncontiguous cartogram}.
\newblock PhD thesis, Virginia Polytechnic Institute, 1989.

\bibitem{MRS10}
W.~Meulemans, A.~van Renssen, and B.~Speckmann.
\newblock Area-preserving subdivision schematization.
\newblock In {\em Geographic Information Science}, volume 6292 of {\em Lecture
  Notes in Computer Science}, pages 160--174, 2010.

\bibitem{Alisa}
A.~Miller.
\newblock {TED} talk: The news about the news.
\newblock
  \url{http://www.ted.com/talks/alisa_miller_shares_the_news_about_the_news},
  2008.

\bibitem{monmonier2014lie}
M.~Monmonier.
\newblock {\em How to lie with maps}.
\newblock 1991.

\bibitem{Task_C}
S.~Nusrat and S.~Kobourov.
\newblock Task taxonomy for cartograms.
\newblock In {\em IEEE Eurographics Conference on Visualization (EuroVis--short
  papers)}, 2015.

\bibitem{cartogram-star}
S.~Nusrat and S.~Kobourov.
\newblock The state of the art in cartograms.
\newblock In {\em IEEE Eurographics Conference on Visualization
  (EuroVis--STARs)}, 2016.

\bibitem{Olson}
J.~Olson.
\newblock Noncontiguous area cartogram.
\newblock {\em The Professional Geographer}, 28(4):371 -- 380, 1976.

\bibitem{Class2}
T.~Pelkofer.
\newblock Using cartograms to learn about {Latin American Demographics}.
\newblock \url{http://laii.unm.edu/outreach/common/retanet/science_math/}.

\bibitem{Raisz34}
E.~Raisz.
\newblock The rectangular statistical cartogram.
\newblock {\em Geographical Review}, 24(3):292--296, 1934.

\bibitem{Hans2}
H.~Rosling.
\newblock {T{ED} Talk: The best stats you've ever seen}.
\newblock
  \url{http://www.ted.com/talks/hans_rosling_shows_the_best_stats_you_ve_ever_seen},
  2006.

\bibitem{RR13}
R.~E. Roth.
\newblock An empirically-derived taxonomy of interaction primitives for
  interactive cartography and geovisualization.
\newblock {\em {IEEE} Transaction on Visualization and Computer Graphics},
  19(12):2356--2365, 2013.

\bibitem{silvia2009people}
P.~J. Silvia and C.~M. Barona.
\newblock Do people prefer curved objects? angularity, expertise, and aesthetic
  preference.
\newblock {\em Empirical studies of the arts}, 27(1):25--42, 2009.

\bibitem{SKI98}
S.~Skiena.
\newblock {\em The algorithm design manual}.
\newblock Springer, 1997.

\bibitem{Stasko06}
J.~T. Stasko.
\newblock Evaluating information visualizations: Issues and opportunities.
\newblock In {\em Presentation at Beyond Time and Errors: Novel Evaluation
  Methods for Visualization, {BELIV'06}}, 2006.

\bibitem{Steven_law}
S.~S. Steven.
\newblock On the psychophysical law.
\newblock {\em Psychological Review}, 64(3), 1957.

\bibitem{Hui}
H.~Sun and Z.~Li.
\newblock Effectiveness of cartogram for the representation of spatial data.
\newblock {\em The Cartographic Journal}, 47(1):12 -- 21, 2010.

\bibitem{Manting}
M.~Tao.
\newblock {\em Using Cartograms in Disease Mapping}.
\newblock PhD thesis, The University of Sheffield, 2010.

\bibitem{teghtsoonian1965judgment}
M.~Teghtsoonian.
\newblock The judgment of size.
\newblock {\em The American journal of psychology}, pages 392--402, 1965.

\bibitem{Tobler73}
W.~Tobler.
\newblock A continuous transformation useful for districting.
\newblock {\em Annals, {NY} Academy of Sciences}, 219:215 -- 220, 1973.

\bibitem{Tobler04}
W.~Tobler.
\newblock Thirty five years of computer cartograms.
\newblock {\em Annals of the Association of American Geographers}, 94:58--73,
  2004.

\bibitem{ks07}
M.~van Kreveld and B.~Speckmann.
\newblock On rectangular cartograms.
\newblock {\em Computational Geometry}, 37(3):175--187, 2007.

\bibitem{Jen}
J.~A. Ware.
\newblock {\em Using animation to improve the communicative aspect of
  cartograms}.
\newblock PhD thesis, Michigan State University, 1998.

\bibitem{Weh93}
S.~Wehrend.
\newblock Appendix {B}: Taxonomy of visualization goals.
\newblock In {\em Visual cues: Practical data visualization.}, pages 203--212,
  1993.

\end{thebibliography}

\begin{IEEEbiography}[{\includegraphics[width=1in,height=1.25in,clip,keepaspectratio]{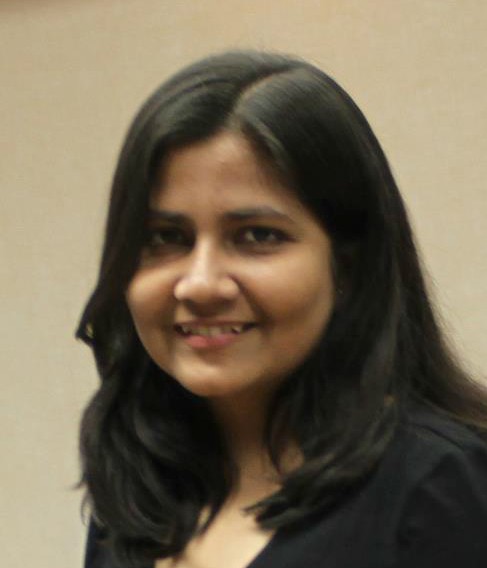}}]
{Sabrina Nusrat}
 is a PhD student at the Department of Computer Science at the University of Arizona. In 2012 she obtained a Masters degree in Computer Science from the University of Saskatchewan. Her current research interest is in visualization and visual analytics, with a focus on geo-referenced visualization. 

\end{IEEEbiography}

\begin{IEEEbiography}[{\includegraphics[width=1in,height=1.25in,clip,keepaspectratio]{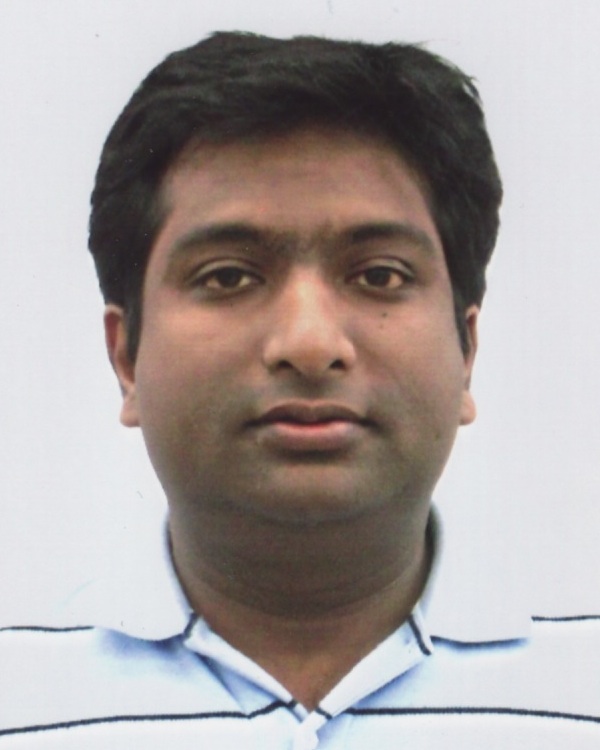}}]
{Md.~Jawaherul Alam} 
is a research scientist at the University of California, Irvine. He received a PhD in Computer Science from the University of Arizona in 2015. His research interests include algorithms for graphs and maps, graph drawing, and information visualization.

\end{IEEEbiography}

\begin{IEEEbiography}[{\includegraphics[width=1in,height=1.25in,clip,keepaspectratio]{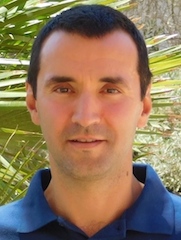}}]
{Stephen Kobourov} is a Professor at the Department of Computer Science at the University of Arizona. He received a BS degree in Mathematics and Computer Science from Dartmouth College and MS and PhD degrees from Johns Hopkins University. His research interests include information visualization, graph theory, and geometric algorithms.

\end{IEEEbiography}
\end{document}